\documentclass[12pt]{iopart}

\usepackage{amsthm}
\usepackage{amsmath}
\usepackage{amssymb}
\usepackage{graphicx}
\usepackage{bm}

\newcommand{\R}{\mathbb{R}}
\newcommand{\C}{\mathbb{C}}
\newcommand{\N}{\mathbb{N}}

\newcommand{\F}{\mathcal{F}}
\newcommand{\A}{\mathcal{A}}
\newcommand{\B}{\mathcal{B}}
\newcommand{\sgn}{\text{sgn}}
\newcommand{\green}{G}
\newcommand{\kernel}{K}

\newcommand{\req}[1]{(\ref{eq:#1})}
\newcommand{\abs}[1]{\left|#1\right|}
\newcommand{\set}[1]{\{ #1 \}}

\newcommand{\fourier}[1]{\F \left\{ #1 \right\}}
\newcommand{\hilbert}[1]{\mathcal{H} \left\{ #1 \right\}}
\newcommand{\ifourier}[1]{\F^{-1} \left\{ #1 \right\}}
\newcommand{\ilaplace}[1]{\mathcal{L}^{-1} \left\{ #1 \right\}}
\newcommand{\timederivative}[1]{\frac{\partial}{\partial t} #1}
\newcommand{\secondtimederivative}[1]{\frac{\partial^2}{\partial t^2} #1}
\newcommand{\ltext}[1]{\qquad \text{#1} \qquad}
\newcommand{\stext}[1]{\quad \text{#1} \quad}

\newcommand{\T}{T}
\newcommand{\x}{\mathbf{x}}

\newcommand{\patt}{p_{\text{att}}}
\newcommand{\tpatt}{\tilde{p}_{\text{att}}}
\newcommand{\qatt}{q_{\text{att}}}
\newcommand{\ta}{\tilde{\alpha}}
\newcommand{\imaginary}{{\bf i}}

\newcommand{\supp}{\mbox{supp}}
\newcommand{\Heavi}{H}

\newtheorem{theorem}{Theorem}[section]
\newtheorem{lemma}{Lemma}[section]

\theoremstyle{definition}
\newtheorem{definition}{Definition}[section]
\newtheorem{example}{Example}[section]
\newtheorem{remark}{Remark}

\def\Xint#1{\mathchoice
  {\XXint\displaystyle\textstyle{#1}}%
  {\XXint\textstyle\scriptstyle{#1}}%
  {\XXint\scriptstyle\scriptscriptstyle{#1}}%
  {\XXint\scriptscriptstyle\scriptscriptstyle{#1}}%
  \!\int}
\def\XXint#1#2#3{{\setbox0=\hbox{$#1{#2#3}{\int}$}
    \vcenter{\hbox{$#2#3$}}\kern-.5\wd0}}

\begin{document}

\title[Causality Analysis \LaTeXe]{Causality Analysis of Frequency Dependent Wave Attenuation}

\author{Richard Kowar}

\address{Department of Mathematics, University of Innsbruck,
Technikerstra\ss{}e 21a/2, A-6020 Innsbruck, Austria}
\ead{richard.kowar@uibk.ac.at}

\author{Otmar Scherzer}

\address{Computational Science Center, University of Vienna,\\
Nordbergstra\ss{}e 19, A-1090 Vienna, Austria and\\ RICAM, Austrian Academy of Sciences, Altenbergerstra\ss{}e 69, A-4040 Linz, Austria}
\ead{otmar.scherzer@univie.ac.at}

\author{Xavier Bonnefond}

\address{Institut de Math\'ematiques de Toulouse, Universit\'e Paul Sabatier, 31062 Toulouse Cedex 9, France}
\ead{bonnefond@mip.ups-tlse.fr}

\begin{abstract}
The work is inspired by thermo-and photoacoustic imaging, where recent efforts are devoted to take into account
attenuation and varying wave speed parameters. In this paper we derive and analyze causal equations describing
propagation of attenuated pressure waves. We also review standard models, like frequency power laws, and the
thermo-viscous equation and show that they lack causality in the parameter range relevant for biological
photoacoustic imaging. To discuss causality in mathematical rigor we use the results and concepts of linear
system theory. We present some numerical experiments, which show the physically unmeaningful behavior of standard
attenuation models, and the realistic behavior of the novel models.
\end{abstract}


\vspace{2pc}
\maketitle

\section{\label{sec:level1} Introduction}

The work is inspired by thermo-and photoacoustic imaging (see e.g. \cite{Tam86,KuWanStoWan04,XuWan06,PatSch07,Wan08} for some articles
related to the subject),
where the problem is the reconstruction of the \emph{absorption density} from measurements of the pressure
outside of the object. This is the \emph{Inverse Problems} according to the forward problem, which maps the
absorption density onto the pressure by solving the \emph{standard} wave equation. Various reconstruction methods
have been suggested in the literature for photoacoustic imaging, which can for instance be found in the excellent survey
\cite{KucKun08}. Recent efforts have been made to take into account attenuation \cite{PatGre06,RivZhaAna06,BurGruHalNusPal07} and varying wave
speed \cite{HriKucNgu08}. The standard model of attenuation (which is reviewed in Section \ref{sec:attenuation}) is
formulated in the frequency range and models the physical reality that high frequency components of waves are attenuated
more rapidly over time.

In this paper we review standard attenuation models, like \emph{power laws} \cite{Ros07,Sza94,Sza95,SusCob04,WatHugBraMil00} and the
\emph{thermo-viscous} wave equation \cite{KinFreCopSan00}. In this context, we discuss causality,
which is the desired feature of attenuation models.
The lack of causality of standard models in the parameter range relevant for photoacoustic imaging requires to
investigate novel equations, which are derived in Section \ref{sec:attenuation} and the following.

We base the derivation of causal attenuation models on the mathematical concept of \emph{linear system theory}, which can for instance be 
found in the book of H\"ormander \cite{Hoe03}.
In Section \ref{sec:waveeq} the abstract formulations are translated in equations
which are formally similar to the wave equation. However, in general, the novel equations are integro-differential equations.
An important issue is that the equations are formulated as inhomogeneous equations with homogeneous initial conditions,
which is not standard for attenuated wave equations, where typically the equations are considered homogeneous with inhomogeneous
initial conditions. For the standard wave equation these two concepts are equivalent, but only the one considered here, is
mathematically sensible for the attenuation model.

The approach leads to some novel causal attenuation models, in particular power law models (valid for a bounded frequency range),
which are documented in the literature to be relevant for biological specimen (in the terminology used later on this means that
$\gamma \in (1,2]$ -- see \cite[Chapter 7]{Web00}) and also for instance also for castor oil, which satisfies a power law with
index $\gamma=1.66$ \cite{Sza94}.
These models are presented in Section \ref{sec:gamgen}.
The rotationally symmetric examples, presented in Section \ref{sec:numerics}, illustrate the unphysical behavior of some existing
attenuation models. Aside from unmeaningly physical effects, the stable and convergent numerical implementation of attenuated,
non-causal wave equations is an unconsidered problem since these equations lack the Courant-Friedrich-Levy (CFL) condition \cite{CouFriLev67}.
The attenuation models considered here have a finite front wave speed and therefore can be implemented in a stable manner.
Thus aside from physical considerations also from a point of view of stable numerical solution of wave equations
questions of causality are most relevant.

Concerning the presentation of the paper, the basic notation and mathematic results are summarized in the appendix.

\section{Linear System Theory}
\label{sec:lin_opt}
This section surveys \emph{linear system theory} (see e.g. \cite{Pap62,Hoe03}), which provides the link between linear systems
and convolution operators. This analysis is essential for the analysis.
For notational convenience, when we speak about functions they are understood in the most wide meaning of the word, and can
for instance be distributions.

In the following, we give a characterization of \emph{causal} functions and operators.
\begin{definition}
\label{def:defA}
\begin{enumerate}
\item A function $f:=f(\x,t)$ defined on the Euclidean space over time (i.e. in $\R^4$) is said to be \emph{causal} if it satisfies
\[
 f(\x,t) = 0 \ltext{for} t < 0\;.
\]
\item In this paper $\A$ (with and without subscripts) denotes a real (that is, it is a mapping between sets of real functions on $\R^4$)
and bounded operator.
\begin{itemize}
\item $\A$ is \emph{translation invariant} if for every function $f$ and every linear transformation $L:=L(\x,t):=(\x-\x_0,t-t_0)$,
with $\x_0 \in \R^3$ and $t_0 \in \R$, it holds that
\[
    \A (f \circ L)  = (\A f)L \;.
\]
Here $\circ$ denotes the decomposition, i.e., $(f \circ L)(\x,t) = f(L(\x,t))$
\item $\A$ is called \emph{causal}, if it maps causal functions to causal functions.
\item
The operator $\A$ has a \emph{causal domain of influence} if the function
\[
\T(\x):=\sup \set{t : \A \delta_{\x,t}(\x,\tau) = 0 \text{ for all } \tau \leq t} \ltext{ for all } \x\in\R^3
\]
is rotationally symmetric and the derivative with respect to the radial component $\T'$ satisfies
\begin{equation}
\label{eq:defT(r)}
    0 < \left(\T'(r)\right)^{-1} \leq  c_0 < \infty\;.
\end{equation}
For convenience of notation we identify $\T(\abs{\x}) = \T(\x)$.

The function $\T$, presumably it exists, corresponds to the travel time of a
wavefront initiated in ${\bf 0}$ at $t=0$. \req{defT(r)} guarantees that the wavefront speed is finite.
\end{itemize}
\end{enumerate}
\end{definition}

\begin{remark}\label{rema:c0finite}
If the operator $\A$ models a physical process in a \emph{homogeneous and isotropic medium}, then $\A$ is shift invariant and
$\A \delta_{\x,t}$ is rotationally symmetric.

If $\T$ exists, and in addition satisfies \req{defT(r)}, then the property of a
causal domain of influence guarantees that a wavefront can propagate with a speed of at most $c_0$.
\end{remark}

Now, we recall a fundamental mathematical theorem (see \cite[Theorem 4.2.1]{Hoe03}) of systems theory, which
relates invariant operators with \emph{space--time} convolutions.

\begin{theorem}\cite[Theorem 4.2.1]{Hoe03}
\label{th:fundamental}
Every linear (causal and) translation invariant operator $\A$ can be written as a space--time convolution operator
with (causal) kernel $\green$. That is, for arbitrary $f$ from a suitable class of functions we have
\begin{equation}
\label{eq:convolution}
   \A f = \green *_{\x,t} f\;.
\end{equation}
\end{theorem}
In analogy to linear system theory we call the kernel $\green$ the \emph{Green function} of $\A$. According to Definition
\ref{def:defA} the considered operators are real and therefore the according Green functions are real valued too. From the definition
of the Green function it follows that
\[
   \green = \A \delta_{\x,t}\;.
\]

In the following example we review the Green function and the convolution operator according to the wave equation.

\begin{example}
We consider the standard wave equation in an isotropic medium with phase speed $c_0\in (0,\infty)$:
\begin{equation}\label{eq:waveeqG0}
   \nabla^2 p -\frac{1}{c_0^2} \secondtimederivative{p} = -f,
\end{equation}
together with \emph{initial conditions}
\begin{equation}
\label{eq:G0init}
\left. p \right|_{t < 0} = 0 \ltext{and} \left. \timederivative{p} \right|_{t < 0} = 0\;.
\end{equation}
With source term $f = \delta_{\x,t}$, the according solution $\green_0$ of \req{waveeqG0} and \req{G0init} is the
\emph{Green function}
\begin{equation}
\label{eq:G0}
    \green_0(\x,t)=\frac{\delta_t \left(t-\frac{\abs{\x}}{c_0}\right)}{4\pi\abs{\x}}\;.
\end{equation}
Because of \req{G0} $\green_0$ is commonly denoted \emph{spherical wave} originating from $\x={\bf 0}$ at time $t=0$.

In the space--frequency domain the Green function can be expressed by
\[
 \fourier{\green_0} := \fourier{\green_0}(\x,\omega) = \frac{1}{\sqrt{2\pi}}
                       \frac{\exp{\left(\imaginary \omega \frac{\abs{\x}}{c_0}\right)}}{4\pi\abs{\x}}\;.
\]
It satisfies
\begin{equation}
\label{eq:g1a}
\nabla\fourier{\green_0}
= \left[\frac{\imaginary \omega}{c_0} -\frac{1}{\abs{\x}}\right] \cdot \fourier{\green_0} \cdot \sgn,
\end{equation}
and is the solution of the \emph{Helmholtz equation}
\begin{equation}
\label{eq:g1b}
  \nabla^2 \fourier{\green_0} + \frac{\omega^2}{c_0^2}\fourier{\green_0}
      = -\frac{1}{\sqrt{2\pi}}\delta_\x\;.
\end{equation}
The operator
\[
  \A_0 f:= \green_0 *_{\x,t} f
\]
is causal and maps a causal function $f$ onto the solution of \req{waveeqG0} and \req{G0init}.
\end{example}

\section{Attenuation}
\label{sec:attenuation}
In the chapter we investigated causality of attenuation models in a homogeneous, isotropic medium.
In mathematical terms, it is common to describe attenuation by a \emph{multiplicative law} in the frequency range:
\begin{definition}
\label{de:III}
A real, bounded, linear, translation invariant operator $\A$ with causal domain of influence is called
\emph{attenuation operator} if there exists a complex function
$\beta_*:=\beta_*(r,\omega)$ such that the associated Green function
$\green:=\A \delta_{\x,t}$ satisfies
\begin{equation}
\label{eq:GG0}
\boxed{
\fourier{\green}(\x,\omega) =  \exp{(-\beta_*(\abs{\x},\omega))}\cdot \fourier{\green_0}(\x,\omega) \ltext{for all} \x\in\R^3,\, \omega\in\R\;.
}
\end{equation}
Here, $\F$ is the Fourier transform (see Appendix).
\end{definition}
We rewrite \req{GG0} in the space--time domain by using
\begin{equation}
\label{eq:defkernel}
     \kernel := \kernel(\x,t) := \frac{1}{\sqrt{2\pi}}\ifourier{\exp{(-\beta_*)}} (\abs{\x},t)\;.
\end{equation}
Therefore
\begin{equation}
\label{eq:repA(f)}
\boxed{
\green (\x,t) = [\kernel *_t \green_0](\x,t)
= \frac{1}{4\pi \abs{\x}} \kernel\left(\x,t-\frac{\abs{\x}}{c_0}\right)\;.}
\end{equation}
Since in the context of this paper the operator $\A$ is real, the associated Green function is real-valued, and consequently
$\beta_*(r,\omega)$ has to be even with respect to $\omega$ (cf. Property \ref{item:even} in Appendix).

\begin{remark}
In physical terms attenuation is a result of frequency dependent energy dissipation
and therefore the ratio of the attenuated and un-attenuated wave amplitude must be smaller or equal to $1$. That is
\[
\exp{(-\Re (\beta_*))} = \abs{\frac{\fourier{\green}}{\fourier{\green_0}}} \leq 1\;.
\]
This implies that the \emph{attenuation coefficient} $\beta_*$ satisfies $\Re(\beta_*)\geq 0$.
\end{remark}

In the literature a special form of the attenuation coefficient is assumed:
\begin{definition}[Standard Form]\label{defstform}
The \emph{standard form} of $\beta_*$ considered in the literature is (see e.g.~\cite{Ros07})
\begin{equation}
\label{eq:defalpha*}
   \beta_*(r,\omega)=\alpha_*(\omega) r \ltext{for} r>0, \,\omega\in\R.
\end{equation}
The function $\alpha := \Re (\alpha_*) $ is called \emph{attenuation law}.
\end{definition}
For the standard form $\beta_*$ several properties for the attenuation operator are at hand. For instance the
following results concerning travel time and causality.
\begin{theorem}\label{rema:propT}
Let $\A$ be an attenuation operator with $\beta_*$ of standard form.
Then the travel time satisfies $\T(\abs{\x})=\abs{\x}/c$ for some constant $0 < c\leq c_0$.
\end{theorem}
\begin{proof}
The definition of the travel time $T$ in Definition~\ref{def:defA} states that $\T(\abs{\x})$ is the largest positive
number such that for the Green function $\green = \A \delta_{\x,t}$
\[
     \green (\x,t) = 0 \ltext{for}  t<\T(\abs{\x})\;.
\]
This condition is equivalent to the condition that the function
\begin{equation}\label{eq:relTG}
  (\x,t) \to \green(\x,t+\T(\abs{\x})) \ltext{ is causal.}
\end{equation}
The operator $\A$ is causal and has causal domain of influence, which implies that $\T(0)=0$ and
$\left(\T'(r)\right)^{-1} \leq  c_0$. Consequently
\begin{equation}\label{eq:relTT'}
     \T(\abs{\x})  = \int_0^{\abs{\x}} T'(r) dr \geq \frac{\abs{\x}}{c_0} \,.
\end{equation}
$\tau(r):= \T(r) - r/c_0$ denotes the largest number such that
$\kernel(\x,\cdot+\tau(\abs{\x})) = \frac{1}{\sqrt{2\pi}} \fourier{\exp (-\beta_*)}$ is causal.
From \req{GG0} it follows that
\[
     \kernel(\x,t) = \kernel(\x/2,t) *_t \kernel(\x/2,t)\,.
\]
This and the Theorem of Supports (cf.~\cite{Hoe03}) imply that
$\tau(r) = 2\tau(r/2)$, and consequently $\tau$ is linear in $r$ and after all $\T$ is linear as well.
\end{proof}
In particular, from \req{repA(f)} and Theorem~\ref{rema:propT} it follows that $\A$ has a causal domain of influence if and only if
$\kernel$ is a causal function.

\begin{remark}
\label{rema:kk}
In the literature (for instance in ~\cite{WatHugBraMil00}) causality is aimed to be enforced by demanding that
\begin{equation}\label{eq:stcauslreq}
    \ifourier{\alpha_*} \ltext{is causal.}
\end{equation}
This is equivalent to that the \emph{Kramers-Kronig} relations for the $m$-th derivative $\alpha_*^{(m)}$ of some function $\alpha_*$
are satisfied, i.e., there exists a non-negative integer $m$, such that
\begin{equation}\label{eq:KKrel}
   \Im( \alpha_*^{(m)} ) = \hilbert{\Re( \alpha_*^{(m)} )}
    = \hilbert{ \alpha^{(m)} } \stext{and}
   \alpha^{(m)}  := \Re( \alpha_*^{(m)}  )
        = -\hilbert{  \Im(\alpha_*^{(m)}) } ,
\end{equation}
where $\hilbert{\cdot}$ is the Hilbert Transform (see Appendix).

\req{stcauslreq} follows already from the causality of $\kernel$:
From the definition of $\kernel$ it follows that
\[
 \abs{\nabla \kernel} = \frac{1}{\sqrt{2\pi}} \abs{\ifourier{\alpha_* \cdot \exp{ (-\alpha_* \abs{\x})}}}\;.
\]
Using some sequence $\set{\x_n}$ with $\x_n \neq {\bf 0}$ and $\x_n \to {\bf 0}$ shows that
\[
  \lim \abs{\nabla \kernel}(\x_n,t) = \frac{1}{\sqrt{2\pi}} \abs{\ifourier{\alpha_*}}(t)\;.
\]
Due to the causality of $\kernel$ the left hand side is zero for $t<0$, and thus $\ifourier{\alpha_*}$ is causal.

However, as we show in Example~\ref{ex:powlaw} below, causality of $\ifourier{\alpha_*}$ does not imply causality of $\kernel$.
In other words, in general, from the causality of $\ifourier{\alpha_*}$ it cannot be deduced that $\A$ has a causal
domain of influence. As a consequence several attenuation models considered in the literature lack causality.
\end{remark}

\begin{example}[Frequency power laws~\cite{Ros07,Sza95}]\label{ex:powlaw}
For the \emph{frequency power law},
\begin{equation}
\label{eq:freqpowlaw}
\boxed{
\alpha(\omega) = \alpha_0\abs{\omega}^\gamma,}
\end{equation}
where $\gamma, \alpha_0\geq 0$ and $\gamma\not\in\N$.
The Kramers-Kronig relation with differentiation index $m=1$ is satisfied for the one-parametric
family of complex extensions (as considered ~\cite{WatHugBraMil00,Sza95})
\begin{equation}\label{eq:al*st}
\boxed{
   \alpha_*(\omega) =
   \frac{\alpha_0}{\cos \left(\frac{\pi}{2} \gamma \right)}(-\imaginary \omega)^\gamma + a_0 \imaginary \omega\;.}
\end{equation}
Indeed \cite[Theorem 7.4.3]{Hoe03} implies that for every polynomial $p$ in $-\imaginary \omega$ with nonegative real
exponents,  $\ifourier{p}$ is causal. Hence if $\gamma>1$, then
$$
    \alpha_*^{II}(\omega):= \alpha_*(\omega) - a_0\,(-\imaginary \omega)\quad a_0\in\R,
$$
has the same real part as $\alpha_*$ and $\ifourier{\alpha_*^{II}}$ is causal. As a consequence the attenuation law $\alpha$
together with the causality condition \req{stcauslreq} does not uniquely
determine the attenuation operator $\A$ (cf. Definition \ref{de:III}).

Let $\alpha_*$ be defined as in \req{al*st}, then according to \cite[Theorem 7.4.3]{Hoe03} $\kernel$, as defined in \req{defkernel},
is causal if and only if $\gamma \in [0,1)$.
Consequently, for frequency power laws with $\gamma > 1$ the according operator $\A$, defined in Definition \ref{de:III}
does {\bf not} have a causal domain of influence.
%We note that for $\gamma > 1$ causality cannot be accomplished by adding to $\alpha_*$ a
%polynomial in $-\imaginary \omega$ with nonegative real exponents $<\gamma$.
%E+-+-
\end{example}

\section{Equations for Attenuated Pressure Waves}
\label{sec:waveeq}

In this section we formulate a causal wave equation which takes into account attenuation and review the literature
(cf.~\cite{Sza94,Sza95,SusCob04,RivZhaAna06,PatGre06}).

Let $\A$ denote a translation invariant operator with causal domain of influence with travel time function $T$ and $c_0$
as in Definition \ref{def:defA}. The Green function $\green$ satisfies \req{repA(f)} and \req{defkernel} and therefore
the according attenuation coefficient is given by
\begin{equation}\label{eq:beta*waveeq}
     \beta_*(\x,\omega)
      =  -\mbox{log} \left\{ 2\sqrt{(2\pi)^3} \abs{\x}
                      \fourier{ \green \left(\x,\cdot+\frac{\abs{\x}}{c_0}\right) }(\omega) \right\}.
\end{equation}
In the following we rewrite the term $\nabla^2 \fourier{\green}$ from which we derive the Helmholtz equation for
$\fourier{\green}$.
Using \req{repA(f)}, which states that $\green = \kernel *_t \green_0 = \A \delta_{\x,t}$,
and the product rule yields
\begin{equation}\label{eq:laplG}
   \nabla^2 \fourier{\green} = \nabla^2 \fourier{\kernel} \cdot \fourier{\green_0} +
                                              2 \nabla \fourier{\kernel} \cdot \nabla \fourier{\green_0}
                                              + \fourier{\kernel} \cdot \nabla^2 \fourier{\green_0}.
\end{equation}
To evaluate this expression, we calculate $\nabla \fourier{\kernel}$ and $\nabla^2 \fourier{\kernel}$.
From \req{defkernel} it follows that
\begin{equation}\label{eq:defg2b}
\begin{aligned}
  \nabla \fourier{\kernel} = -\beta_*' \cdot \fourier{\kernel} \cdot \sgn \,,
\end{aligned}
\end{equation}
where $\beta_*'$ denotes the derivative of $\beta_*(r,\omega)$ with respect to $r$.
This together with the formula \req{der_sgn} in the Appendix implies that
\begin{equation}\label{eq:defg2c}
\begin{aligned}
\nabla^2 \fourier{\kernel}
         =&- \nabla \cdot \left( \beta_*' \cdot \fourier{\kernel} \cdot \sgn \right)\\
         =& - (\nabla \cdot \sgn) \cdot \beta_*' \cdot \fourier{\kernel}
            - (\sgn \cdot \nabla \beta_*') \cdot \fourier{\kernel}
            - (\sgn \cdot \nabla \fourier{\kernel}) \cdot \beta_*' \\
         =& \left[ -\frac{2}{\abs{\x}} \cdot\beta_*'
                   - \beta_*''
                   +  \left(\beta_*'\right)^2\right] \cdot \fourier{\kernel}.
\end{aligned}
\end{equation}
Inserting \req{defg2b} and \req{defg2c} into \req{laplG} and using the identity $\green = \kernel *_t \green_0$ (cf. \req{repA(f)}), shows that
\begin{equation}\label{eq:laplG2}
 \begin{aligned}
   \nabla^2 \fourier{\green}
      =& \left[ -\frac{2}{\abs{\x}} \cdot \beta_*' - \beta_*'' +  \left(\beta_*'\right)^2\right]\cdot\fourier{\green}\\
       &- 2 \beta_*' \cdot \fourier{\kernel} \cdot (\sgn \cdot \nabla \fourier{\green_0}) + \fourier{\kernel} \cdot \nabla^2 \fourier{\green_0}.
 \end{aligned}
\end{equation}
Together with \req{g1a} and \req{g1b}, the last identity simplifies to
\begin{equation}\label{eq:laplG3}
 \begin{aligned}
   \nabla^2 \fourier{\green}
      =& \left[ -\frac{2}{\abs{\x}} \cdot \beta_*'
                   - \beta_*''
                   +  \left(\beta_*'\right)^2\right] \cdot \fourier{\green}
       - 2 \left[\frac{\imaginary \omega}{c_0} -\frac{1}{\abs{\x}}\right] \cdot
                   \beta_*' \cdot \fourier{\green}\\
       &-\frac{\omega^2}{c_0^2} \cdot \fourier{\green}
                          -\frac{1}{\sqrt{2\pi}} \cdot \fourier{\kernel} \cdot \delta_\x\;.
 \end{aligned}
\end{equation}
Since $\fourier{\kernel}(\x,\omega) \cdot \delta_\x = \fourier{\kernel}({\bf 0},\omega) \cdot \delta_\x$, we obtain from \req{laplG3}
the Helmholtz equation
\begin{equation}\label{eq:helmholtz}
\boxed{
\begin{aligned}
~& \nabla^2 \fourier{\green}
       -\left[\beta_*' + \frac{(-\imaginary \omega)}{c_0}\right]^2 \cdot\fourier{\green}\\
 =& -\beta_*'' \cdot \fourier{\green}
           -\frac{1}{\sqrt{2\pi}} \fourier{\kernel}({\bf 0},\omega)\cdot \delta_\x\\
 =& -\beta_*'' \cdot \fourier{\green}
           -\frac{1}{2\pi} \exp{(-\beta_*({\bf 0},\omega))} \cdot \delta_\x\,.
\end{aligned}
}
\end{equation}
To reformulate \req{helmholtz} in space--time coordinates, we introduce two convolution operators:
\begin{equation}\label{eq:defD*D*'}
   D_*f := \kernel_* *_{\x,t} f \ltext{and}
   D_*'f :=\kernel_*' *_{\x,t} f ,
\end{equation}
where the kernels $\kernel_*$ and $\kernel_*'$ are given by
\begin{equation}
\label{eq:defKK'}
   \kernel_*:=\kernel_*(\x,t) := \kernel_*(\abs{\x},t) \ltext{and}
   \kernel_*(r,t)  := \frac{1}{\sqrt{2\pi}} \ifourier{\beta_*'}(r,t)
\end{equation}
and
\begin{equation}
\label{eq:defKK'2}
  \kernel_*':=\kernel_*'(\x,t):=
  \kernel_*'(\abs{\x},t)
\ltext{and}
  \kernel_*'(r,t)  = \frac{1}{\sqrt{2\pi}} \ifourier{\beta_*''}(r,t).
\end{equation}
Using these operators and applying the inverse Fourier transform to \req{helmholtz} gives
\begin{equation}\label{eq:waveeq+}
\boxed{
  \nabla^2 \green
       -\left[D_*
         + \frac{1}{c_0} \timederivative{}\right]^2  \green
                  =  -D_*'  \green
                     - \kernel({\bf 0},t) \delta_\x.}
\end{equation}
For a general source term $f$, $\patt:=\A f = G*_{\x,t} f$ solves the equation
\begin{equation}
\label{eq:waveeq+2}
\boxed{
     \nabla^2 \patt -\frac{1}{c_0^2} \secondtimederivative{\patt} =  - \A_s f\;,}
\end{equation}
where $\A_s$ denotes the space--time convolution operator with kernel
\begin{equation}\label{eq:defKs}
\begin{aligned}
   \kernel_s:=\kernel_s(\x,t) :=  -\B \green
           +D_*' \green
          + \kernel({\bf 0},t) \cdot \delta_\x
\end{aligned}
\end{equation}
where
\begin{equation}\label{eq:B}
    \B:= D_*^2 + \frac{2}{c_0} D_*\timederivative{}\;.
\end{equation}
Equation~\req{waveeq+2} is the pressure wave equation that obeys attenuation with attenuation coefficient~\req{beta*waveeq}.

\begin{remark}\label{rema:alpha*2}
In this remark we consider again the standard model, as in Definition~\ref{defstform}.
For this case the wave equation~\req{waveeq+2} can be casted in a form that resembles the standard attenuation wave equation (cf. Example~\ref{ex:stwaveeq}).
Since $\kernel$ is causal, it follows that $\kernel_*$ is causal too (the argumentation is analogous to Remark~\ref{rema:kk})
and therefore the operator $D_*$ is well-defined for all causal functions.
Moreover, since $\kernel_*'=0$, it follows that $D_*'\equiv 0$. Using that $\kernel_*$ depends only on $t$ it follows that
\[
       (D_* \green) *_{\x,t} f =  [\kernel_* *_t \green] *_{\x,t} f
       =  \kernel_* *_t [\green *_{\x,t} f] = D_*(\green *_{\x,t} f).
\]
Convolving each term in \req{waveeq+} with a function $f$, using the previous identity and that $D_*' \equiv 0$, it follows that
\begin{equation} \label{eq:waveeq+3}
  \nabla^2 \patt
       -\left[D_* + \frac{1}{c_0} \timederivative{}\right]^2  \patt = - f \;.
\end{equation}
\end{remark}

In the following we review some wave equations obeying attenuation, which are frequently considered in the
literature:

\begin{example}\label{ex:stwaveeq}
\begin{itemize}
\item For $\gamma>0$ and $\gamma \not\in \N$, denote by $D_t^\gamma$ be the \emph{Riemann-Liouville fractional derivative}
(see \cite{KilSriTru06,Pod99})) with respect to time. It is defined in the Fourier domain by
\begin{equation}\label{eq:defDtga}
        \fourier{D_t^{\gamma}f} = (-\imaginary \omega)^\gamma \fourier{f}
\end{equation}
and satisfies
\begin{equation}\label{eq:propDtgamma}
    D_t^{2\gamma}f = D_t^\gamma D_t^{\gamma}f \ltext{and}
    \timederivative{D_t^{\gamma}f} = D_t^\gamma \timederivative{f} = D_t^{\gamma+1}f.
\end{equation}
Now, we consider the attenuation coefficient
\begin{equation}\label{eq:betapowlaw}
     \beta_*(r,\omega):= \ta_0 (-\imaginary \omega)^\gamma r  \ltext{ with }  \ta_0:=\alpha_0/\cos(\gamma \pi/2)\,,
\end{equation}
which satisfies the attenuation law
\[
\Re (\beta_*)(r,\omega)= \alpha (\omega) r \ltext{ and } \alpha(\omega)=\alpha_0\abs{\omega}^\gamma
\]
(cf. Example~\ref{ex:powlaw} and~\cite{WatHugBraMil00,Sza95}).
Let $D_*$ denote the time-convolution operator with kernel $K_*$ defined by ~\req{defKK'} and  ~\req{betapowlaw}. Then
form \req{defDtga} and  $ K_* = \ifourier{\ta_0(-\imaginary \omega)^\gamma}/\sqrt{2 \pi} $ it follows that $D_* = \ta_0 D_t^\gamma$.
In \cite{SusCob04,RivZhaAna06} (see also~\cite{Sza94,Sza95}) the following equation for the pressure function $\patt$
of attenuated waves is investigated:
\begin{equation}
\label{eq:standwaveeq}
\boxed{
\begin{aligned}
  \nabla^2 \patt
       &-\left[\ta_0 D_t^\gamma
         + \frac{1}{c_0} \timederivative{}\right]^2  \patt
      = -f  \;,
\end{aligned}}
\end{equation}
which is equivalent to equation \req{waveeq+3} with operator $D_* = \ta_0 D_t^\gamma$.
Let $\A$ denote the solution operator of \req{standwaveeq}, then from \cite[Theorem 7.4.3]{Hoe03} it follows that $\A$ has a causal
domain of influence only for $\gamma \in [0,1)$.
\item Let $\gamma>0,\gamma\not\in\N$.
      Neglecting in \req{standwaveeq} the operator $\ta_0^2 D_t^{2\gamma}$ (which one finds after expanding the decomposition operator)
      one finds Szabo's equation \cite{Sza94}
\begin{equation}\label{eq:waveeqszabo}
\begin{aligned}
    \nabla^2 \patt -\frac{1}{c_0^2} \secondtimederivative{\patt}
       - \frac{2\ta_0}{c_0} D_t^{\gamma+1}(\patt)
            = -f\;.
\end{aligned}
\end{equation}
This equation is equivalent to equation \req{waveeq+3} if we define the kernel of $D_*$ by \req{defKK'} with
\begin{equation}\label{eq:alpha*szabo}
\begin{aligned}
\beta_*(r,\omega):= \ta_0 (-\imaginary \omega)^\gamma r  \stext{ and }
  \alpha_*(\omega) =  \frac{\imaginary \omega}{c_0}
                      + \frac{1}{c_0} \sqrt{\left(-\imaginary \omega\right)^2
                    + 2\ta_0 c_0 (-\imaginary \omega)^{\gamma+1}}\;.
\end{aligned}
\end{equation}
Again, if $\A$ denotes the solution operator of \req{waveeqszabo}, then \cite[Theorem 7.4.3]{Hoe03} implies that $\A$ has a causal
domain of influence only for $\gamma\in [0,1)$.
\end{itemize}
In the literature, the standard attenuation models \req{standwaveeq} and \req{waveeqszabo} are considered as
\emph{homogeneous Cauchy problems} with \emph{inhomogeneous} initial conditions. In contrast, in our setting, we consider
\emph{inhomogeneous Cauchy problems} with \emph{homogeneous} initial conditions. In the following section we show that the two
concepts can be equivalent. However, in general, only the concept suggested here leads to a rigorous
framework, in which we can define solution operators for attenuated wave equations.
\end{example}

For the readers convenience, we summarize some important notation and facts in the following table.
Note the difference between $\kernel$, $\kernel_*$ and $\kernel_*'$, respectively, with respect to the involved exponential function.
\begin{center}
\begin{tabular}{|c|c|c||c|}
\hline
Kernel & General & Standard Form & Convolution Operator\\
\hline
$\kernel$ \req{defkernel} & $\frac{1}{\sqrt{2\pi}} \ifourier{ \exp{(-\beta_*)}}$
& $\frac{1}{\sqrt{2\pi}} \ifourier{\exp{(-\alpha_*  \abs{\x})}}$ & \\
$\kernel_*$ \req{defKK'} & $\frac{1}{\sqrt{2\pi}} \ifourier{\beta_*'}$ & $\frac{1}{\sqrt{2\pi}} \ifourier{\alpha_*}$ & $D_*$ \req{defD*D*'}\\
$\kernel_*'$ \req{defKK'}& $\frac{1}{\sqrt{2\pi}} \ifourier{\beta_*''}$ & 0 & $D_*'$ \req{defD*D*'}\\
\hline
\end{tabular}
\end{center}

\section{The homogeneous Cauchy problem with memory}
\label{sec:relfinit}

We consider the standard attenuation model $\beta_*(r,\omega)=\alpha_*(\omega)r$.
Let $\A$ denote a translation invariant operator with causal domain of influence and and let the operator $D_*$ be
as defined as in \req{defD*D*'}.

In this section we investigate under which conditions the inhomogeneous wave equation \req{waveeq+3} with homogeneous initial conditions \req{G0init} (where $p$ is replaced by $\patt$) and
the homogeneous equation
\begin{equation} \label{eq:eqpatt2}
   \nabla^2 \qatt
       -\left[D_*
         + \frac{1}{c_0} \timederivative{}\right]^2  \qatt = 0
\end{equation}
with the inhomogeneous initial condition
\begin{equation}\label{eq:eqpatt2init}
   \qatt = q_0  \ltext{for} t \leq 0 \ltext{and}
\left. \timederivative{\qatt} \right|_{t=0+} = \left. \timederivative{q_0} \right|_{t=0-}
\end{equation}
are equivalent. That is, both equations have the same solution for $t > 0$.

\begin{theorem}
Assume that \req{eqpatt2}, \req{eqpatt2init} and \req{waveeq+3}, \req{G0init} have unique solutions, respectively.
Then $\qatt=\patt$ for $t>0$
if and only if $q_0$ and $\patt$ are related by the following conditions
\begin{equation}\label{eq:cauchycond1}
\begin{aligned}
        \varphi:=\lim_{t\to 0-} q_0 = \lim_{t\to 0+} \patt\,,  \qquad
        \psi:=\lim_{t\to 0-} \timederivative{q_0} = \lim_{t\to 0+} \timederivative{\patt}
\end{aligned}
\end{equation}
and
\begin{equation}\label{eq:cauchycond2}
\boxed{
\begin{aligned}
        \Heavi \cdot \B q_0 = -f + \frac{1}{c_0^2} \left(\psi \cdot \delta_t + \varphi \cdot \timederivative{\delta_t} \right) \,,
\end{aligned}}
\end{equation}
with $\B$ is as in \req{B} and $\Heavi$ is the Heaviside function.
\end{theorem}

\begin{proof}
\begin{description}
 \item{$\Rightarrow$:}
Assume that $\qatt=\patt$ for $t>0$. Then, using that $\patt = 0$ for $t<0$, implies that
\begin{equation}\label{eq:qpq0}
\patt = \Heavi \cdot \qatt   \ltext{and}    \qatt = \patt + q_0   \,.
\end{equation}
In particular, property \req{cauchycond1} holds. Moreover, \req{qpq0} implies
\begin{equation}\label{eq:relpattqatt}
  \nabla^2 \patt = \Heavi \cdot \nabla^2 \qatt \ltext{and}
  \secondtimederivative{\patt} = \Heavi \cdot \secondtimederivative{\qatt}
                 + \psi \cdot \delta_t  + \varphi \cdot \timederivative{\delta_t}.
\end{equation}
Since
\begin{equation}
\label{eq:propB}
  \B + \frac{1}{c_0^2} \secondtimederivative{\;}
     =  \left[D_* + \frac{1}{c_0} \timederivative{}\right]^2
\end{equation}
it follows from \req{relpattqatt}, \req{propB} and \req{qpq0} that
\begin{equation} \label{eq:eqHqaatt2}
\begin{aligned}
   \nabla^2 \patt&   -\B\patt - \frac{1}{c_0^2} \secondtimederivative{\;}\patt  +f
         = \Heavi \cdot \left[\nabla^2 \qatt -\B \qatt - \frac{1}{c_0^2} \secondtimederivative{\;}\qatt \right] \\
             &-\B\patt + \Heavi \cdot \B\qatt
              -\frac{1}{c_0^2} \left(\psi \cdot \delta_t + \varphi \cdot \timederivative{\delta_t} \right)
               + f  \,.
\end{aligned}
\end{equation}
Using the definitions of $\qatt$ and $\patt$, \req{eqHqaatt2} simplifies to
\[
    -\B\patt + \Heavi \cdot \B\qatt
          -\frac{1}{c_0^2} \left(\psi \cdot \delta_t + \varphi \cdot \timederivative{\delta_t} \right)
               + f = 0\,.
\]
Since $\B$ is a causal operator and $\patt$ is a causal function, we have $\B\patt= \Heavi \cdot \B\patt$.
This together with \req{qpq0} implies that $-\B\patt + \Heavi \cdot \B\qatt  = \Heavi \cdot \B q_0$.
Hence
\[
     \Heavi \cdot \B q_0
     -\frac{1}{c_0^2} \left(\psi \cdot \delta_t + \varphi \cdot \timederivative{\delta_t} \right)
      + f = 0\,
\]
and thus \req{cauchycond2} holds. This proves the first direction of the theorem. \\
\item{$\Leftarrow$:}
To prove the opposite direction let
\[
    \tpatt := \Heavi\cdot\qatt \ltext{such that} \qatt =  \tpatt + q_0\,.
\]
We prove that $\patt = \tpatt$ holds for $t>0$. Similarly as in part a) of the proof it follows that
\[
\begin{aligned}
   \nabla^2 \tpatt&   -\B\tpatt - \frac{1}{c_0^2} \secondtimederivative{\;}\tpatt
         = \Heavi \cdot \left[\nabla^2 \qatt -\B\qatt - \frac{1}{c_0^2} \secondtimederivative{\;}\qatt \right] \\
             &+ \Heavi \cdot \B q_0
              -\frac{1}{c_0^2} \left(\psi \cdot \delta_t + \varphi \cdot \timederivative{\delta_t} \right)
\end{aligned}
\]
holds. Since $\qatt$ solve problem \req{eqpatt2}, \req{eqpatt2init} and condition \req{cauchycond2} is satisfied, the last identity simplifies to
\[
\begin{aligned}
   \nabla^2 \tpatt&   -\B\tpatt - \frac{1}{c_0^2} \secondtimederivative{\;}\tpatt
            = -\tilde f\,.
\end{aligned}
\]
Hence we have shown that  $\tpatt$ solves problem \req{waveeq+3}, \req{G0init} and since this problem has the
unique solution $\patt$, it follows $\tpatt=\patt$. In summary we have shown that
$$
          \patt = \tpatt = \qatt \ltext{ for }t>0\,,
$$
which proves the assertion.
\end{description}
\end{proof}

\begin{remark}
In the absence of attenuation the operator $\B$ is the zero operator and condition \req{cauchycond2} reduces to
\[
 f =  \frac{1}{c_0^2} \left(\psi \cdot \delta_t + \varphi \cdot \timederivative{\delta_t} \right) \,.
\]
In this case the solutions of
\[
\boxed{
 \begin{array}{lcl}
 \nabla^2 p -\frac{1}{c_0^2} \secondtimederivative{p} = -f\,, \qquad & ~\qquad \qquad & \nabla^2 q -\frac{1}{c_0^2} \secondtimederivative{q} = 0\,,\\
 \left. p \right|_{t < 0} = 0\,, \left. \timederivative{p} \right|_{t < 0} = 0\,, &  &
 \left. q \right|_{t = 0} = \varphi \,, \left. \timederivative{q} \right|_{t = 0} = \psi ,
 \end{array}}
\]
are identical for $t>0$.
\end{remark}

\section{The thermo-viscous wave equation}
\label{sec:thvcase}

In this section we show that the \emph{thermo-viscous} wave equation (see e.g.~\cite{KinFreCopSan00})
is not causal (see Theorem \ref{le:kk_tv} below).
The formalism introduced here will enable us to derive a causal variant of the thermo-viscous equation
which satisfies the same attenuation law.

The thermo-viscous wave equation models propagation of pressure waves in viscous media and reads as follows
\begin{equation}\label{eq:thvwaveeq}
\begin{aligned}
\boxed{
    \left(I+\tau_0 \timederivative{\;} \right)\nabla^2 \patt
       -\frac{1}{c_0^2}\secondtimederivative{\patt}
            = -F\;.}
\end{aligned}
\end{equation}
Here $\tau_0$ and $c_0$ denotes the \emph{relaxation time} and the \emph{thermodynamic speed},
respectively and $F$ models sources.

In the following we transform the thermo-viscous wave equation into the form \req{waveeq+}, which enables us to deduce that
the thermo-viscous equation is not causal. For these purpose we consider the attenuation coefficient
\begin{equation}\label{eq:beta0}
\beta_*(r,\omega) = \alpha_*(\omega) r \ltext{with}
\alpha_*(\omega) = \frac{\imaginary \omega}{c_0} - \frac{\imaginary \omega}{c_0 \sqrt{1-\tau_0 \imaginary \omega}}\,.
\end{equation}
and the time convolution operators $T^{1/2}$ and $L^{1/2}$ with kernels
\[
\kernel_{T^{1/2}} := \frac{1}{\sqrt{2 \pi}} \ifourier{ (1-\imaginary \tau_0\omega)^{-1/2}}
\ltext{and}
\kernel_{L^{1/2}} := \frac{1}{\sqrt{2 \pi}}\ifourier{(1-\tau_0 \imaginary \omega)^{1/2}}\,,
\]
respectively. Since $\kernel_*$ satisfies \req{defKK'} it can be rewritten in the following form
\begin{equation}\label{eq:K*tv}
\kernel_* = \frac{1}{\sqrt{2\pi}} \ifourier{\frac{\imaginary \omega}{c_0} - \frac{\imaginary \omega}{c_0\,\sqrt{1-\imaginary \omega \tau_0}} }
          = -\frac{1}{c_0} \timederivative{\delta_t} + \frac{1}{c_0}\timederivative{\kernel_{T^{1/2}}}\;.
\end{equation}
Therefore the according convolution operator $D_*$ is given by
\begin{equation}\label{eq:D*AB}
    D_* = -\frac{1}{c_0}\timederivative{\;} + \frac{1}{c_0} T^{1/2} \timederivative{\;}\;.
\end{equation}
In the following we summarize some properties of the operators $T^{1/2}$, $L^{1/2}$ and $D_*$, and the associated kernels.

\begin{lemma}\label{lem:thv}
The kernel functions $\kernel_{T^{1/2}}$, $\kernel_{L^{1/2}}$ and the operators $L^{1/2}$, $T^{1/2}$, $D_*$ respectively, satisfy:
\begin{enumerate}
\item\label{item:no} For $\tau_0=0$, $T^{1/2}= L^{1/2} = I$ and $\kernel_{T^{1/2}}=\kernel_{L^{1/2}}=\delta_t$.
\item
\begin{equation}\label{eq:defkT}
          \kernel_{T^{1/2}}(t) = \frac{\sqrt{2\pi} \Heavi(t) \exp{(-t/\tau_0)}}{\Gamma(1/2)\tau_0^{1/2} t^{1/2}}\;.
\end{equation}
\item $L^{1/2}$ is the inverse of $T^{1/2}$,
\item  Let $ L:=\left( L^{1/2} \right)^2 $ and $ T:=\left( T^{1/2} \right)^2$, then
\begin{equation}
\label{eq:LT}
   L = I+\tau_0\timederivative{} \ltext{and} T = L^{-1}\;.
\end{equation}
\item \label{item:Dstern}
      $D_*' \equiv 0$ and for $\tau_0 = 0$ also $D_* \equiv 0$.
\item \label{item:D*T}
      \[
      \begin{aligned}
      \left[ D_* + \frac{1}{c_0}\timederivative{\;}\right]^2 = \frac{1}{c_0^2} T\,\secondtimederivative{\;}
\end{aligned}
\]

\end{enumerate}
\end{lemma}

\begin{proof}
\begin{enumerate}
\item The first item is a trivial consequence of properties of the Fourier transform $\ifourier{\cdot}$.
\item
With the substitution $s=-\imaginary \omega \tau_0$ we derive the relation with the inverse Laplace transformation $\ilaplace{\cdot}$
(for a definition and some basic properties see the appendix of this paper)
\begin{equation}
\begin{aligned}
\ifourier {(1-\imaginary \omega \tau_0)^{-1/2}}(t)
 &= \frac{1}{\imaginary \tau_0\sqrt{2\pi} } \int_{-\imaginary \infty}^{\imaginary \infty} \exp{(s t /\tau_0)}\cdot (1+s)^{-1/2} d s \\
 &= \frac{\sqrt{2\pi}}{\tau_0}\ilaplace{(1+s)^{-1/2}}(t/\tau_0).
\end{aligned}
\end{equation}
Using the properties \req{pr1} and \req{pr2} of the inverse Laplace transformation the assertion follows.
\item From
$$
    K_{T^{1/2}} *_t K_{L^{1/2}} = K_{L^{1/2}} *_t K_{T^{1/2}} =\frac{1}{\sqrt{2\,\pi}}\,\ifourier{1} = \delta_t\,,
$$
it follows that for each function $f$
\begin{equation}
\begin{aligned}
         T^{1/2} \,L^{1/2}\,f  &=   K_{T^{1/2}} *_t K_{L^{1/2}} *_t f = \delta_t*_t f = f  \\
         L^{1/2}\,T^{1/2}\,f  &=   K_{T^{1/2}} *_t K_{L^{1/2}} *_t f = \delta_t*_t f = f\;.
\end{aligned}
\end{equation}
\item Since
      \[
      \kernel_{L^{1/2}} *_t  \kernel_{L^{1/2}} = \frac{1}{\sqrt{2\pi}}\ifourier{1-\imaginary \omega \tau_0}
      = \delta_t -\tau_0\timederivative{\delta_t},
      \]
      it follows that
$$
        L\,f = \kernel_{L^{1/2}} *_t  \kernel_{L^{1/2}} *_t f
             = \left(\delta_t -\tau_0\timederivative{\delta_t}\right) *_t f
             = \left(I+\tau_0\timederivative{\;}\right)\,f \,.
$$
The assertion $T= L^{-1}$ is then a consequence of the previous item.
\item Since $\kernel_*$ does not depend on $\abs{\x}$ and $\kernel_*'$ is the kernel of $D_*'$, it follows that $\kernel_*'=0$, i.e. $D_*'\equiv0$.
The second statement is a direct consequence of Item \ref{item:no} which states that $T^{1/2}=I$ for $\tau_0=0$.
\item Follows from \req{D*AB}.
\end{enumerate}
\end{proof}

The thermo-viscous wave equation \req{thvwaveeq} can be put in formal relation to the wave equation \req{waveeq+} by identifying
an appropriate operator $D_*$ as in~\req{D*AB}:

Utilizing Item~\ref{item:D*T} of Lemma~\ref{lem:thv} in equation \req{waveeq+3} and taking into account that $D_*'\equiv 0$ (cf.
Item \ref{item:Dstern} of Lemma~\ref{lem:thv}) shows that the solution of the thermo-viscous wave equation \req{thvwaveeq} with
$F:=L\,f$ satisfies
\[
\begin{aligned}
 \nabla^2 \patt - \left[ D_* + \frac{1}{c_0}\timederivative{\;}\right]^2\patt =
\nabla^2 \patt - \frac{1}{c_0^2} T\,\secondtimederivative{\patt} = -f\;.
\end{aligned}
\]
Conversely, the solution of equation \req{waveeq+3} with $D_*$ defined as in~\req{D*AB} satisfies
the thermo-viscous wave equation \req{thvwaveeq} with $F=L\,f$.

\begin{theorem}
\label{le:kk_tv}
Let $\alpha_*$ be defined as in \req{beta0}.
Then $\Re(\alpha_*)$ and $\Im(\alpha_*)$ satisfy the Kramers-Kronig relation, but
the solution operator $\A$ of the thermo-viscous wave equation does not have a causal domain of influence.
\end{theorem}

\begin{proof}
Since $K_*$ defined as in~\req{K*tv} is causal, it follows that  $\Re(\alpha_*)$ and $\Im(\alpha_*)$ satisfy the Kramers-Kronig relation.
From \cite[Theorem 7.4.3]{Hoe03} it follows that the  kernel $\kernel :=  \frac{1}{\sqrt{2 \pi}} \ifourier{\exp{(-\alpha_* \abs{\x})}}$
is not causal and as a consequence the according solution operator of the thermo--viscous wave equation does not have a causal domain of
influence.
\end{proof}

\begin{remark}
From \req{beta0} it follows that the attenuation law $\alpha=\Re(\alpha_*)$ approximates for small frequencies the frequency power law
with $\gamma=2$.
\end{remark}

\section{A causal thermo-viscous wave equation}
Below we discus a causal variant of the thermo-viscous wave equation.

Let $\alpha_1 \geq 0$. Theorem \ref{le:kk_tv} below shows that the attenuation operator
with attenuation coefficient of standard form $\beta_*(r,\omega)=\alpha_*^c(\omega) r$ and
\begin{equation}\label{eq:beta0caus}
      \alpha_*^c(\omega) = - \frac{\alpha_1 \imaginary \omega}{c_0 \sqrt{1-\tau_0 \imaginary \omega}}\,
\end{equation}
has a causal domain of influence. The operator $D_*$ and its kernel $\kernel_*$ read as follows
\begin{equation}\label{eq:D*ABmod}
    \boxed{D_* := \frac{\alpha_1}{c_0} T^{1/2}\timederivative{\;}   \ltext{ and }
    \kernel_* = \frac{\alpha_1}{c_0} \timederivative{\kernel_{T^{1/2}}}\,.}
\end{equation}
Note that $D_*'\equiv 0$, since $\kernel_*$ does not depend on $\abs{\x}$.

For $\alpha_1=0$, $D_*\equiv 0$ and thus equation \req{waveeq+} with operator $D_*$ defined by \req{D*ABmod}
is the standard wave equation (without attenuation). Since
\[
\left(D_* + \frac{1}{c_0} \timederivative{\;}\right)^2
= \frac{1}{c_0^2} \left[ I+ \alpha_1 T^{1/2}  \right]^2\,
       \secondtimederivative{\;} \ltext{and} L=T^{-1}\,,
\]
\req{waveeq+} can be rewritten as
\begin{equation}\label{eq:cthvmodel}
\boxed{
    \left(I+\tau_0 \timederivative{\;} \right)\nabla^2 \patt
       -\left[\alpha_1 I+L^{1/2}\right]^2\frac{1}{c_0^2} \secondtimederivative{\patt}
            = - \left(I+\tau_0 \timederivative{\;}\right)\,\delta_{\x,t}.}
\end{equation}
%The solution operator of this equation has a causal domain of influence, and satisfies the same attenuation law as
%the thermo-viscous wave equation, which is a small frequency approximation of the frequency power law with $\gamma=2$.

\begin{theorem}
Let $\alpha_*$ and $\alpha_*^c$ be defined as in \req{beta0} and \req{beta0caus}, respectively. Then  $\Re(\alpha_*^c)$ and $\Im(\alpha_*^c)$
satisfy the Kramers-Kronig relation and $\Re(\alpha_*)=\Re(\alpha_*^c)$.
The solution operator $\A$ of equation \req{cthvmodel} has a causal domain of influence.
\end{theorem}

\begin{proof}
Since $K_*$ defined as in~\req{D*ABmod} is causal, it follows that  $\Re(\alpha_*^c)$ and $\Im(\alpha_*^c)$ satisfy the Kramers-Kronig relation.
Comparison of  $\alpha_*$ defined as in \req{beta0} and $\alpha_*^c$ defined as in \req{beta0caus} shows that  $\Re(\alpha_*)=\Re(\alpha_*^c)$.
From \cite[Theorem 7.4.3]{Hoe03} it follows that the  kernel $\kernel :=  \frac{1}{\sqrt{2 \pi}} \ifourier{\exp{(-\alpha_*^c \abs{\x})}}$
is causal and as a consequence the solution operator of equation \req{cthvmodel} has a causal domain of influence.
\end{proof}

\begin{remark}
In ultrasound imaging soft tissue is often modeled as a viscous fluid and therefore \req{cthvmodel} is a
potential model, on which thermoacoustic tomography can be based on.
Moreover, the attenuation of tissue is frequently modeled as a power frequency law with $\gamma\in (1,2)$.
\end{remark}

\section{Causal Wave Equations satisfying Frequency Power Laws for small frequencies with $\gamma \in (1,2]$}
\label{sec:gamgen}
In Example~\ref{ex:powlaw} we have shown that the frequency power law does not yield to a causal wave equation when $\gamma \geq 1$.
In this section we derive causal wave equations for attenuation laws which approximate frequency power laws
\emph{for small frequencies} with exponent $\gamma \in (1,2]$, where for $\gamma=2$ we get the causal variant of the
thermo-viscous wave equation \req{cthvmodel}.

Here we follow the notation of the previous section and introduce the following families of operators:
For constants $\gamma \in (1,2]$, $\tau_0\geq 0$ and $\alpha_1\geq 0$ let $T_\gamma^{1/2}$ and
$L^{1/2}_\gamma$ denote time convolution operators with kernels:
\[
\begin{aligned}
  \kernel_{T_\gamma^{1/2}} &:= \frac{1}{\sqrt{2\pi}}
        \fourier{ \left( 1+(-\imaginary \omega \tau_0)^{\gamma-1} \right)^{-1/2} },\\
  \kernel_{L_\gamma^{1/2}} &:= \frac{1}{\sqrt{2\pi}}
        \fourier{ \left(1+(-\imaginary \omega \tau_0)^{\gamma-1} \right)^{1/2}} .
\end{aligned}
\]
We set $T_\gamma:=\left(T_\gamma^{1/2}\right)^2$ and $L_\gamma:=\left(L_\gamma^{1/2}\right)^2$.
We emphasize that $T_2^{1/2}=T^{1/2}$, where $T^{1/2}$ is the operator in the thermo-viscous case.
The operators $T_\gamma^{1/2}$ and $L_\gamma^{1/2}$ satisfy similar properties as the operators $T^{1/2}$ and $L^{1/2}$
in the thermo-viscous case:

The following lemma is proven analogously as Lemma~\ref{lem:thv}.

\begin{lemma}
\begin{itemize}
\item For $\tau=0$ we have
      \[
      T_\gamma^{1/2}= L_\gamma^{1/2} = I \ltext{and} \kernel_{T_\gamma^{1/2}}=\kernel_{L_\gamma^{1/2}}=\delta_t\;.
      \]
\item $L^{1/2}_\gamma$ is the inverse of $T_\gamma^{1/2}$.
\item Let $D_t^{\gamma-1}$ be the fractional derivative of order $\gamma-1$, as defined as in \req{defDtga}, then
      \[
         L_\gamma=I+\tau_0^{\gamma-1}D_t^{\gamma-1} \,.
      \]
\end{itemize}
\end{lemma}

In analogy to Section \ref{sec:thvcase} we consider now the standard attenuation coefficient with
\begin{equation}\label{eq:beta0causgam}
      \alpha_*(\omega) =   - \frac{\alpha_1 \imaginary \omega}{c_0\sqrt{1+(-\imaginary \tau_0\omega)^{\gamma-1}}}\,.
\end{equation}
Here $\alpha_1$, $\tau_0$ and $c_0$ are positive constants that are medium specific.
The operator $D_*$ and its kernel $\kernel_*$ are given by
\begin{equation}\label{eq:D*ABmodgam}
 \boxed{
     D_* := \frac{\alpha_1}{c_0} T_\gamma^{1/2} \timederivative{\;}
       \ltext{ and }
    \kernel_* = \frac{\alpha_1}{c_0} \timederivative{\kernel_{T_\gamma^{1/2}}}\,.}
\end{equation}
Moroever, the kernel $\kernel$, defined by\req{defkernel}, reads as follows
\begin{equation}\label{eq:gtissue}
\begin{aligned}
    \kernel(\x,t) =   \frac{1}{\sqrt{2\pi}}
                \fourier{\exp\left(
                               \frac{\alpha_1 \imaginary \omega \abs{\x}}
                                    {c_0 \sqrt{1+(-\imaginary \tau_0\omega)^{\gamma-1}}}\right)}(t)
\end{aligned}
\end{equation}
For $\omega$ small we have
\[
\begin{aligned}
   \alpha(\omega)
      \sim \alpha_0  \abs{\tau_0 \omega}^\gamma
\ltext{with}
\alpha_0 = \frac{\sin(\frac{\pi}{2} (\gamma-1))}{2 \tau_0 c_0}.
\end{aligned}
\]
The wave equation \req{waveeq+} with $D_*$ as in \req{D*ABmodgam}  reads as follows
\begin{equation}\label{eq:cthvmodelgam}
   \left(I+\tau_0^{\gamma-1}  D_t^{\gamma-1} \right)\nabla^2 \patt
   -\left[\alpha_1\,I + L^{1/2}_\gamma \right]^2
    \frac{1}{c_0^2} \secondtimederivative{\patt} = -F.
\end{equation}
In particular, for $\gamma=2$ we recover the causal variant of the thermo-viscous wave equation \req{cthvmodel}.

\begin{theorem}
The solution operator of equation \req{cthvmodelgam} has a causal domain of influence.
\end{theorem}
\begin{proof}
From \cite[Theorem 7.4.3]{Hoe03} it follows that $\kernel$ from~\req{gtissue} is causal and thus the solution
operator of \req{cthvmodelgam} has a causal domain of influence.
\end{proof}

\section{Examples}
\label{sec:numerics}

In this section we present some calculations, highlighting the effects of non-causality.
In all examples $\beta_*$ is of standard form \req{defalpha*} and the solution operator $\A$ determined by
$\beta_*$ has the Green function
\[
   \green(\x,t) = \frac{1}{\sqrt{2\,\pi}} \int_\R
             \frac{ \exp{ \left(-\beta_* \right)} \cdot
                    \exp{\left(-\imaginary \cdot (t-\frac{\abs{\x}}{c_0}) \right)}
                  } {4\,\pi \abs{\x}} d\omega\,.
\]
We recall that the operator $\A$ has a causal domain of influence if and only if $\ifourier{\exp{ \left(-\beta_*\right)}}$ is a causal function.
In other words, non-causality can be observed if
\[
     \ifourier{\exp{\left(-\beta_*\right)}}(t) \not=0  \ltext{for some} t<0 \;.
\]
All numerical simulations were performed in MATLAB with the fft-subroutine.

\begin{description}
\item{Frequency power law:}  Let $\alpha = \alpha_0 \abs{\omega}^\gamma$ with some $\gamma > 0$.
The extension $\alpha_*$ by the Kramers-Kroenig relation is given by \req{al*st}.
Fig.~\ref{fig:powlaw} shows simulations of $\ifourier{\exp{\left(-\beta_* \right)}}$, which illustrates that causality only
holds for $\gamma\in [0,1)$.
\begin{figure}[!ht]
\begin{center}
\includegraphics[height=4.0cm,angle=0]{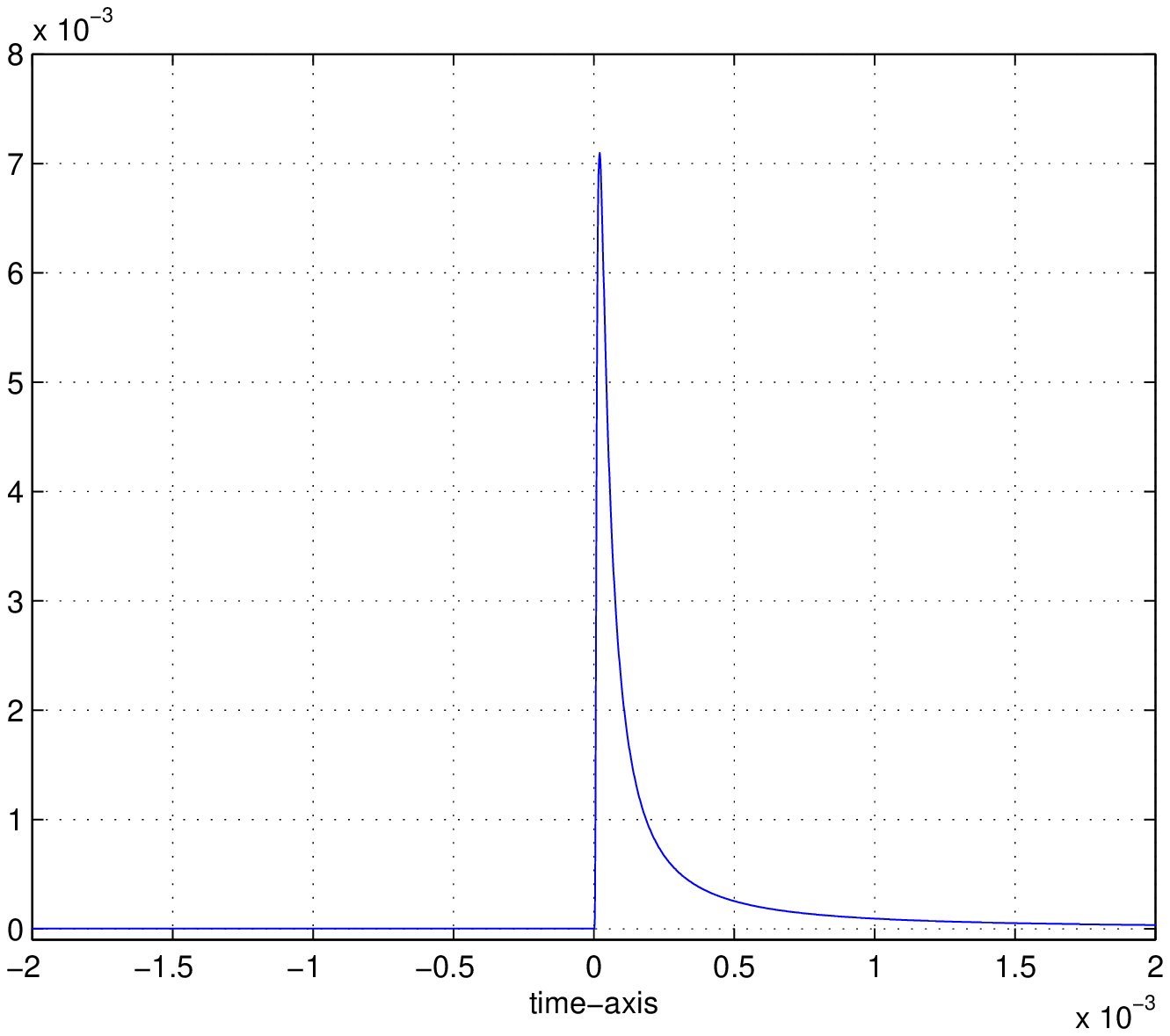}
\includegraphics[height=4.0cm,angle=0]{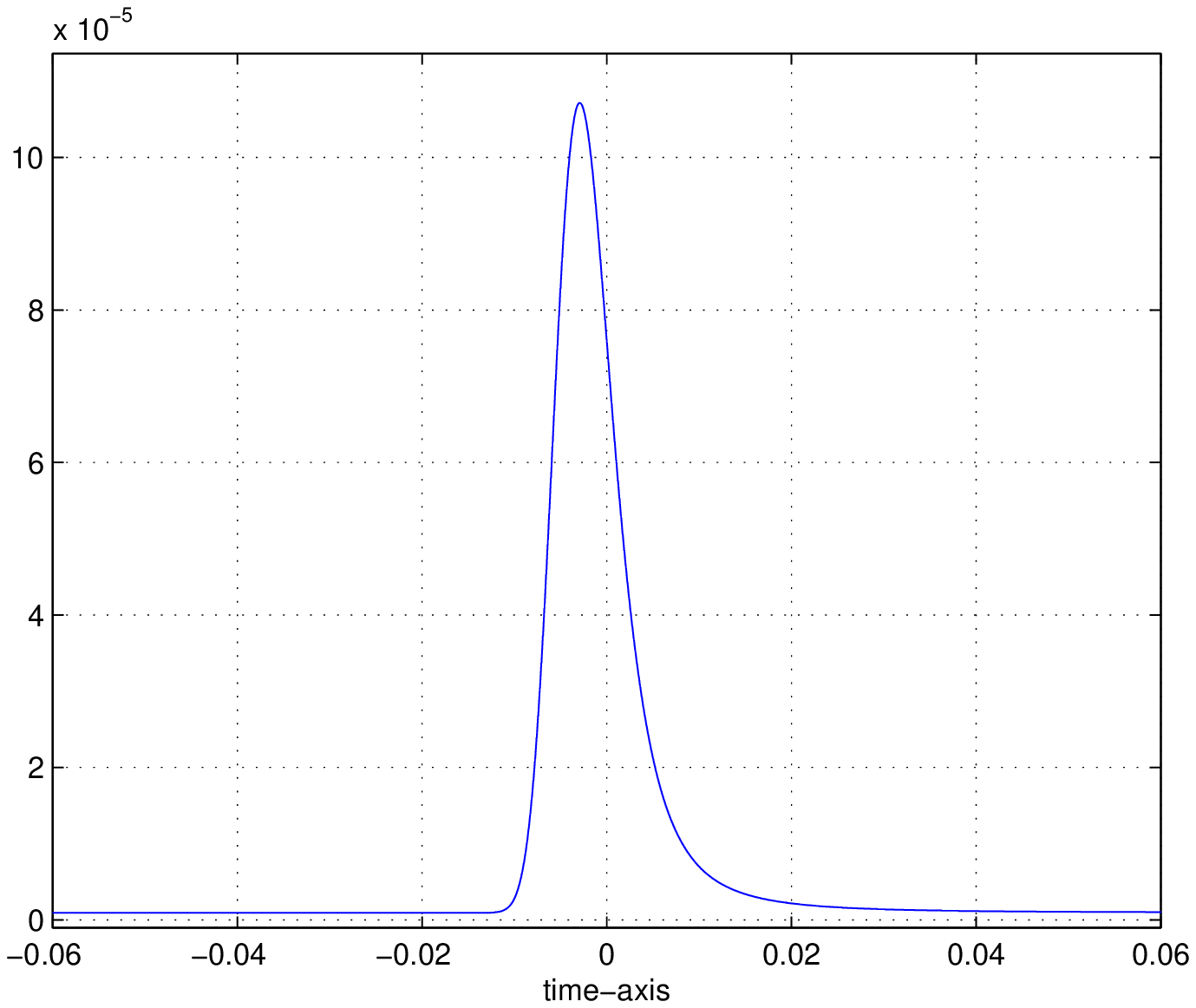}\\
\includegraphics[height=4.0cm,angle=0]{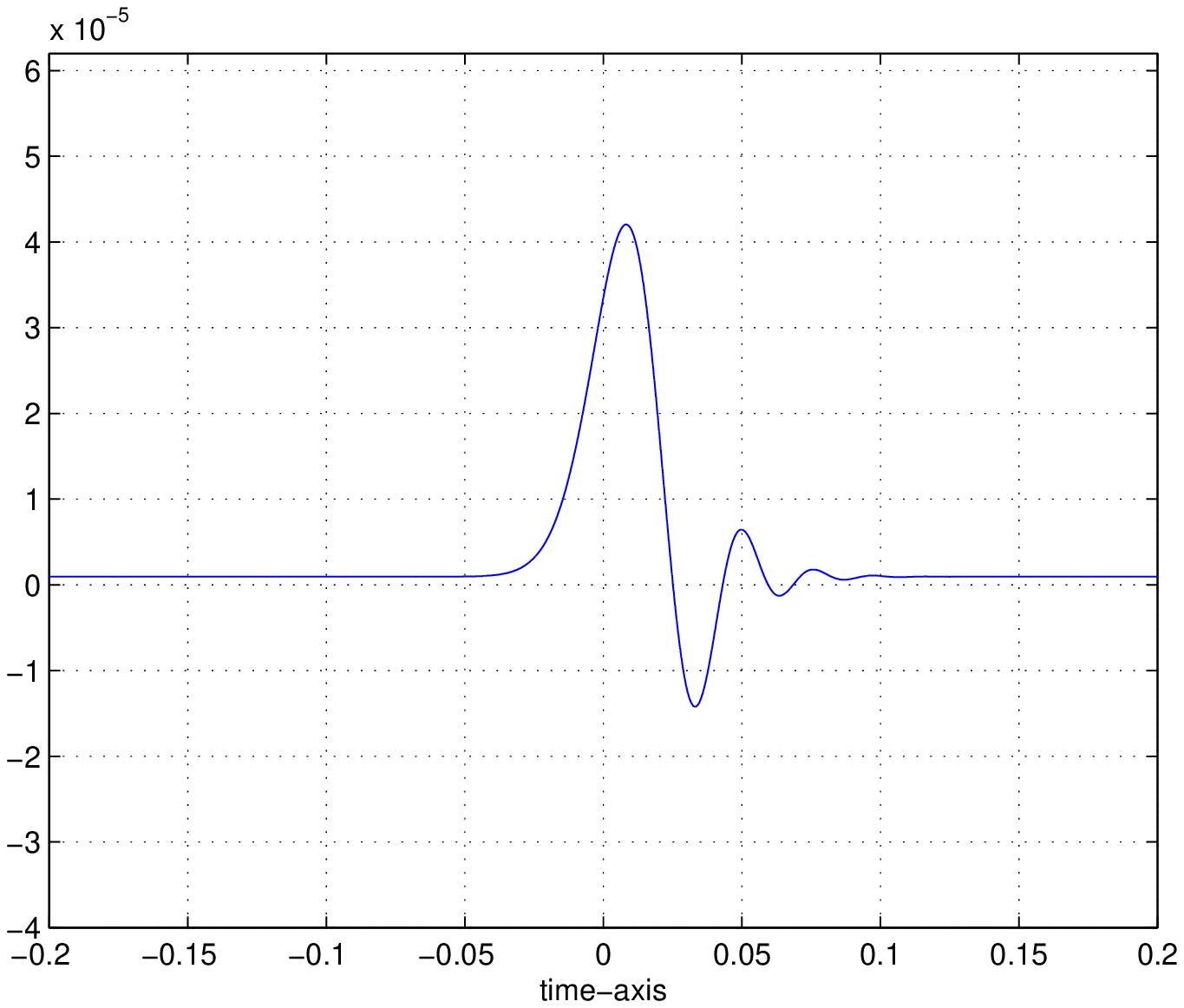}
\includegraphics[height=4.0cm,angle=0]{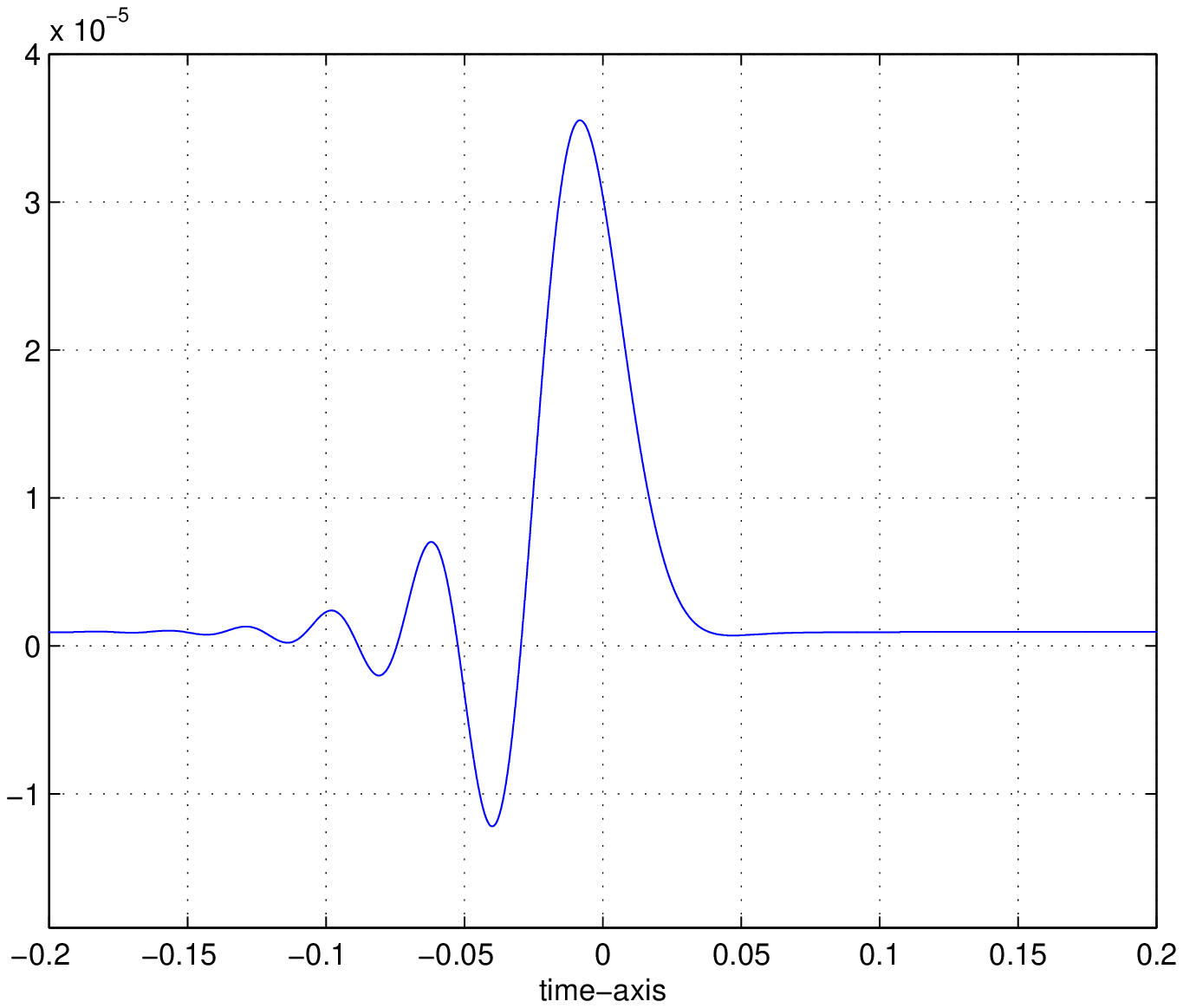}
\end{center}
\caption{Simulation of $\ifourier{\exp{ \left(-\beta_*(\abs{\x},\omega)\right) }}$ for
the frequency power law with $(\gamma,\alpha_0)\in \{(0.5,0.1581),\,(1.5,0.0316),\,(2.7,0.0071),\,(3.3,0.0027)\}$, $c_0=1$ and
$\abs{\x}=\frac{1}{4}$. In the first example $\gamma < 1$ and thus the function is causal. For all other cases it is non causal as predicted by the theory.}
\label{fig:powlaw}
\end{figure}

\item{Szabos's model:} Here $\alpha_*(\omega)$ is as in \req{alpha*szabo}.
In Fig.~\ref{fig:szabo} we show simulations of $\ifourier{\exp{ \left(-\beta_*\right)}}$.
The numerical result confirm the mathematical considerations that causality only holds for $\gamma\in [0,1)$.
\begin{figure}[!hb]
\begin{center}
\includegraphics[height=4.0cm,angle=0]{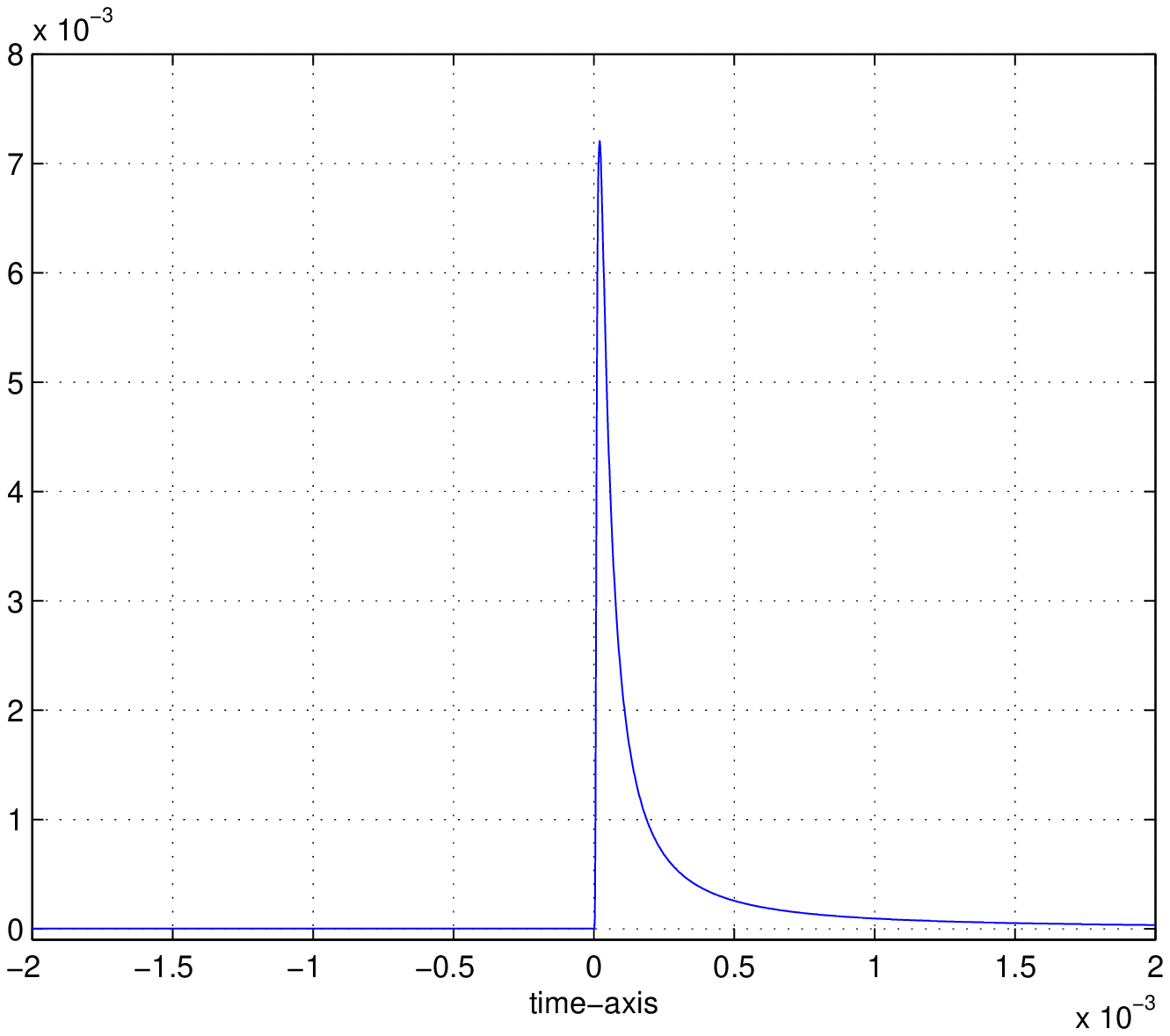}
\includegraphics[height=4.0cm,angle=0]{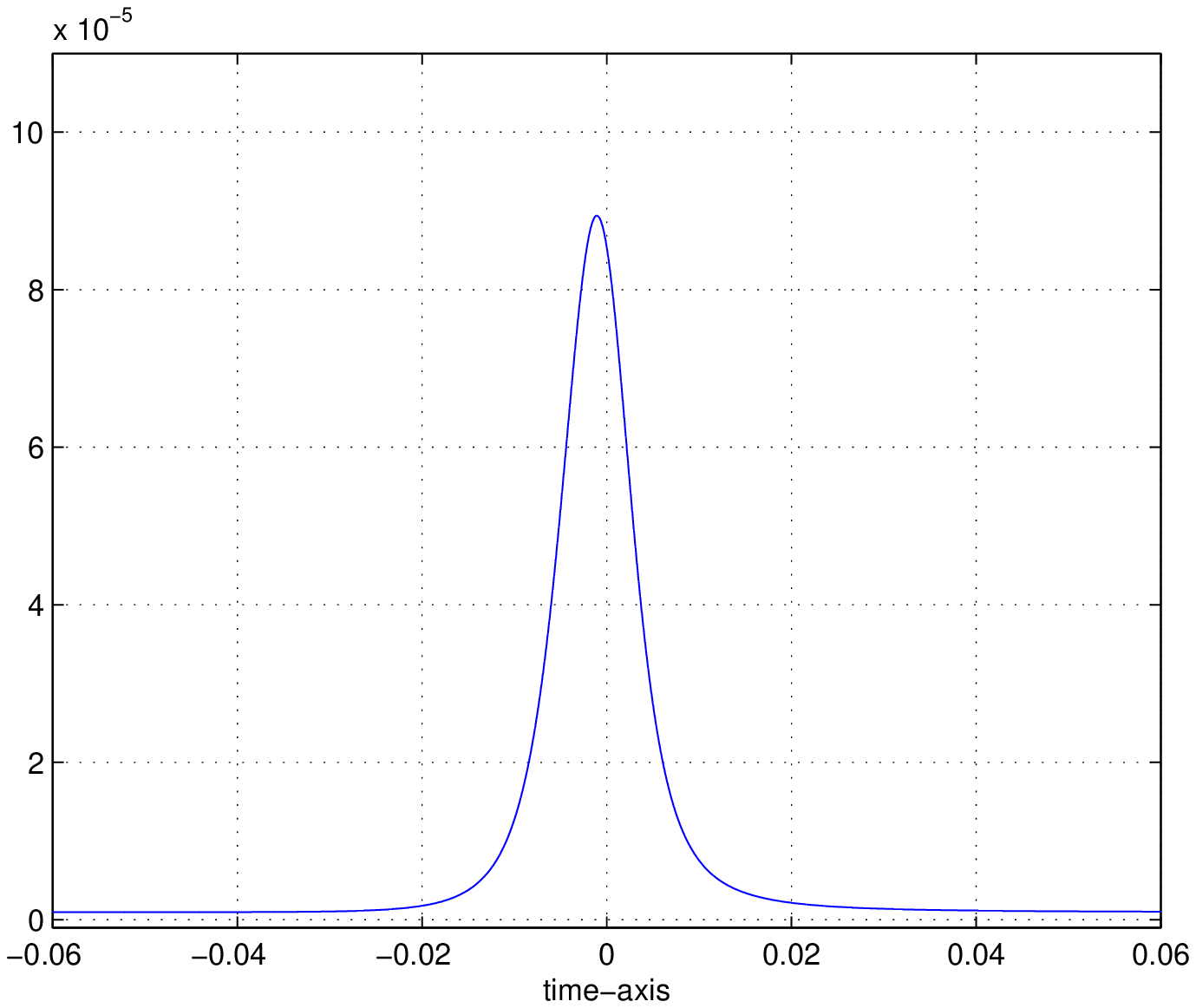}\\
\includegraphics[height=4.0cm,angle=0]{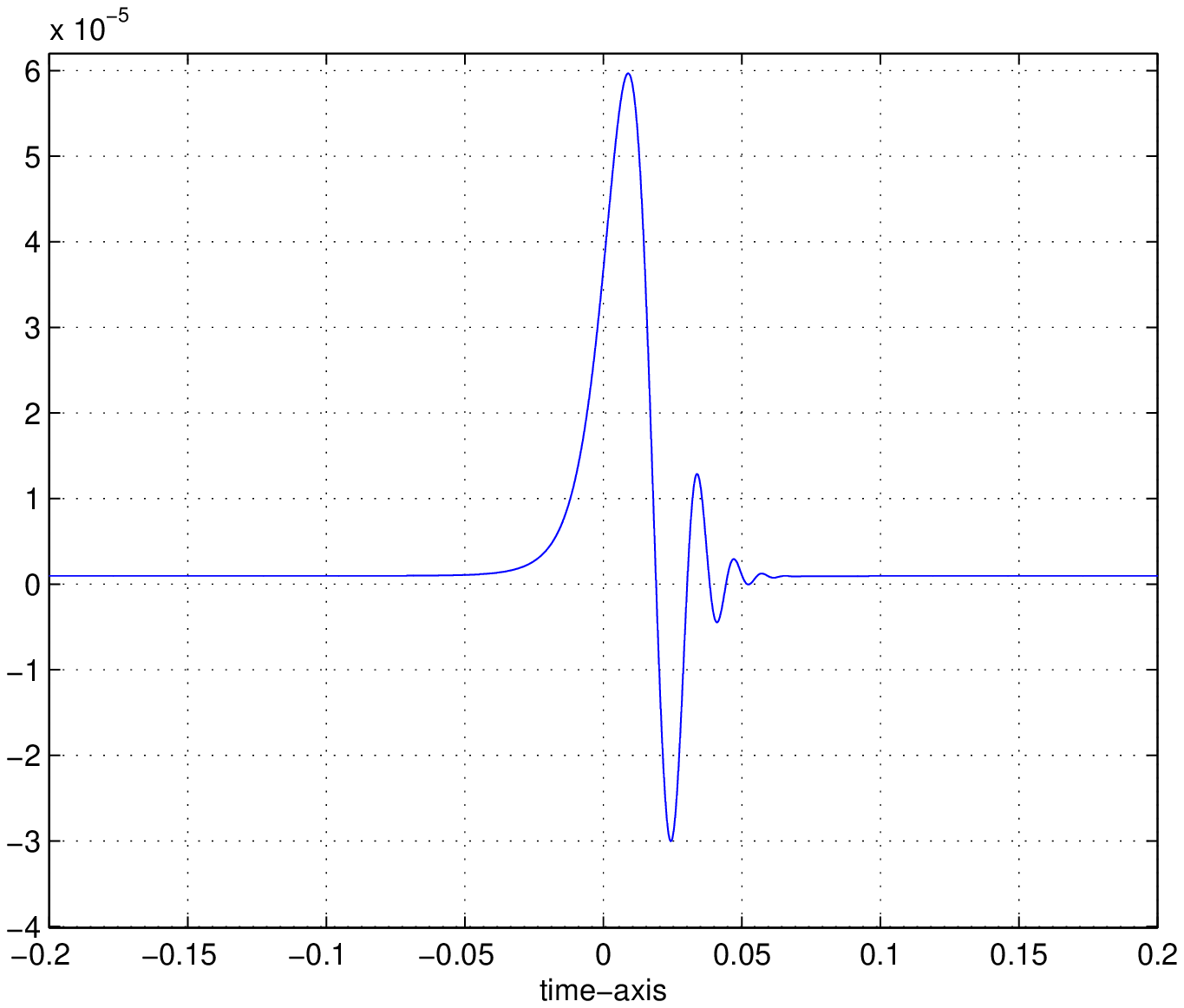}
\includegraphics[height=4.0cm,angle=0]{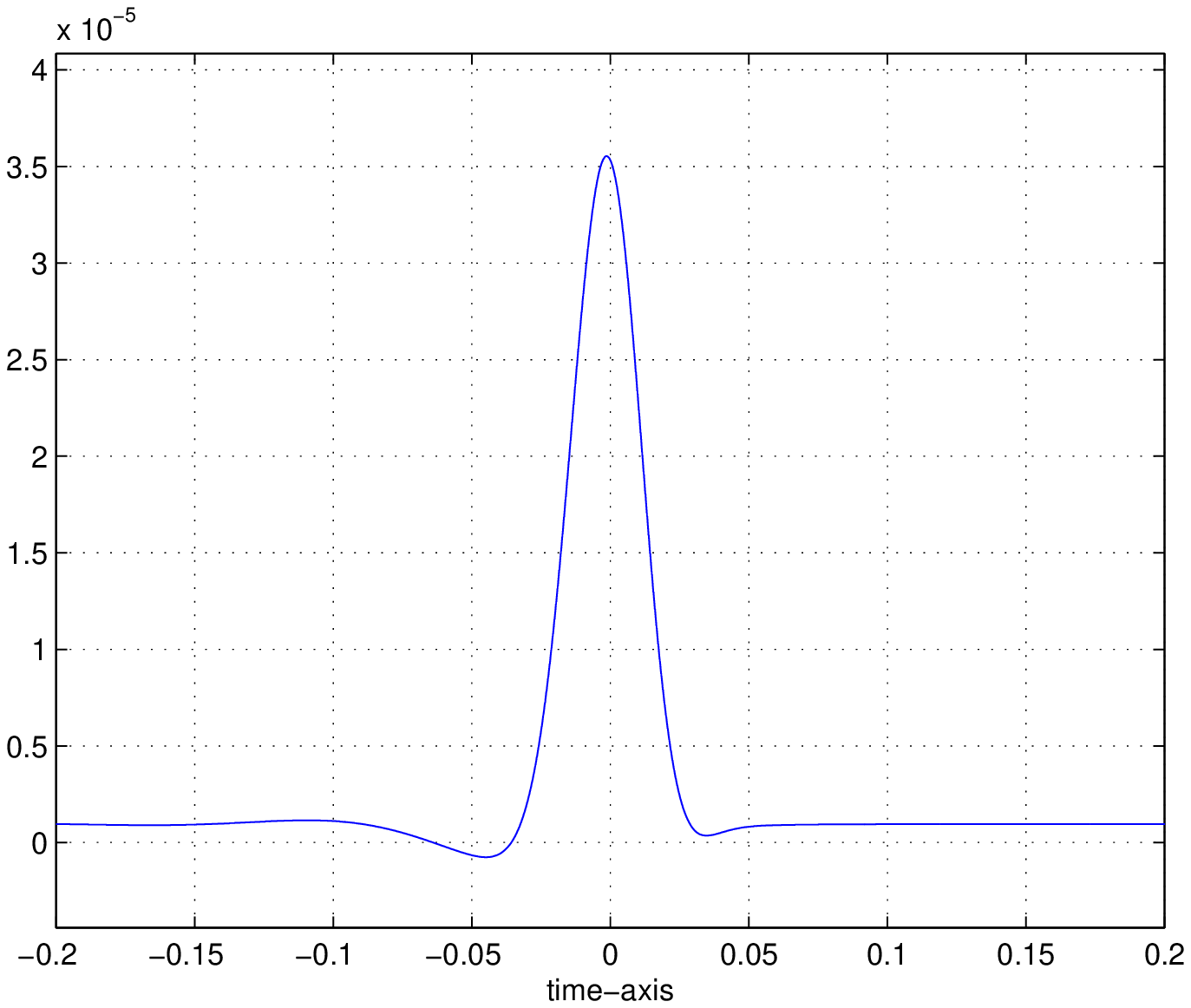}
\end{center}
\caption{Simulation of $\ifourier{\exp{ \left(-\beta_*(\abs{\x},\omega)\right)}}$ for
Szabo's frequency law with $(\gamma,\alpha_0)\in \set{(0.5,0.1581),\,(1.5,0.0316),\,(2.7,0.0071),\,(3.3,0.0027)}$, $c_0=1$ and
$\abs{\x}=\frac{1}{4}$. }
\label{fig:szabo}
\end{figure}

\item{Thermo-viscous wave equation:} There $\alpha_*$ is as in \req{beta0}.
The left pictures in Fig.~\ref{fig:thviscous} shows a simulation of $\ifourier{\exp{\left(-\beta_*\right)}}$ for the
thermo-viscous wave equation \req{thvwaveeq}.
Note that according to \req{beta0} and \req{beta0caus}  the attenuation laws of the thermo-viscous wave equation and the causal
variant \req{cthvmodel} differ just by a multiplicative constant $\alpha_1$.
A simulation of $\ifourier{\exp{\left(-\beta_*\right)}}$ with $\alpha_1=1$ for the causal variant \req{cthvmodel}
of the thermo-viscous wave equation is shown in the right pictures of Fig.~\ref{fig:thviscous}
\begin{figure}[!ht]
\begin{center}
\includegraphics[height=4.0cm,angle=0]{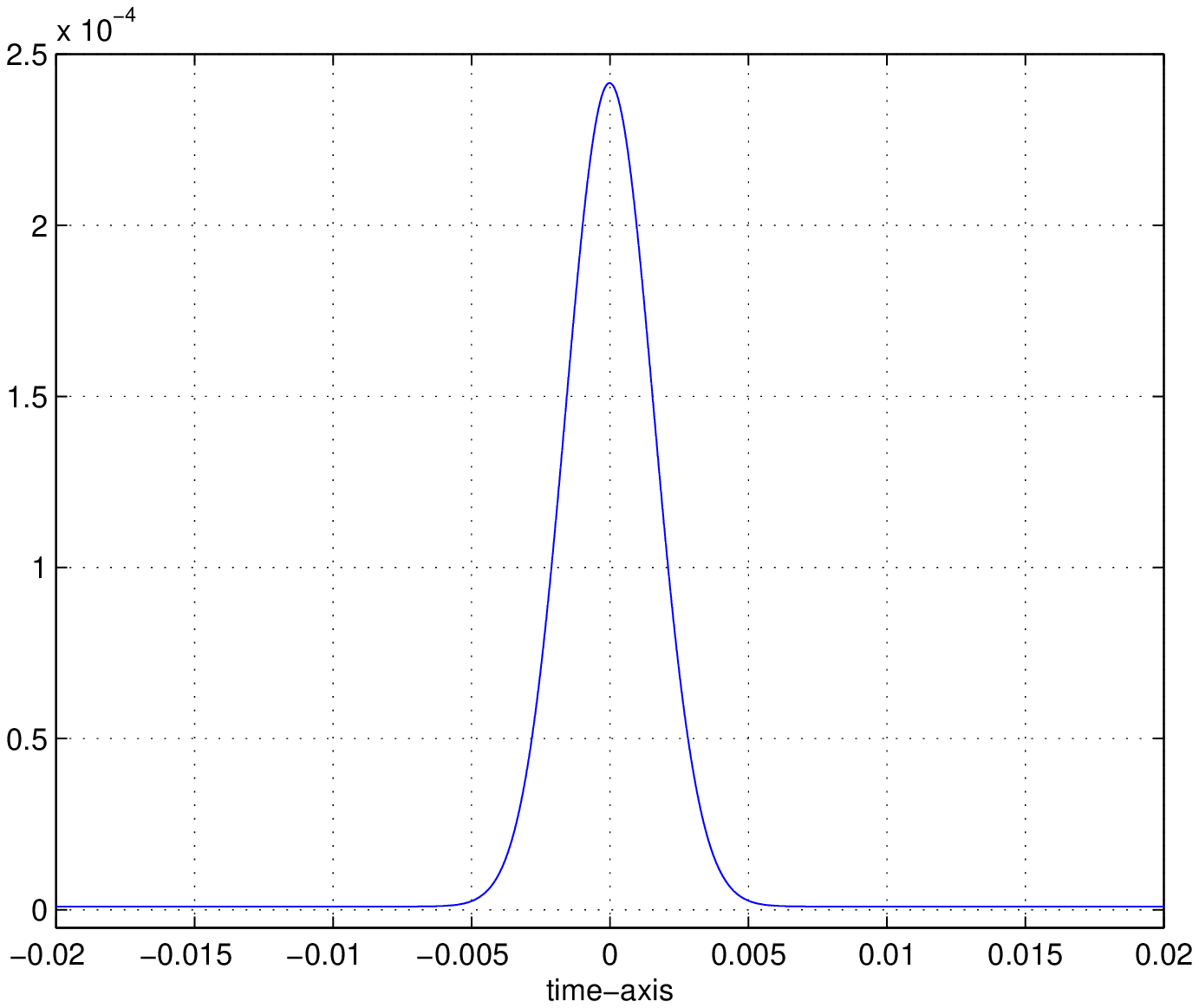}
\includegraphics[height=4.0cm,angle=0]{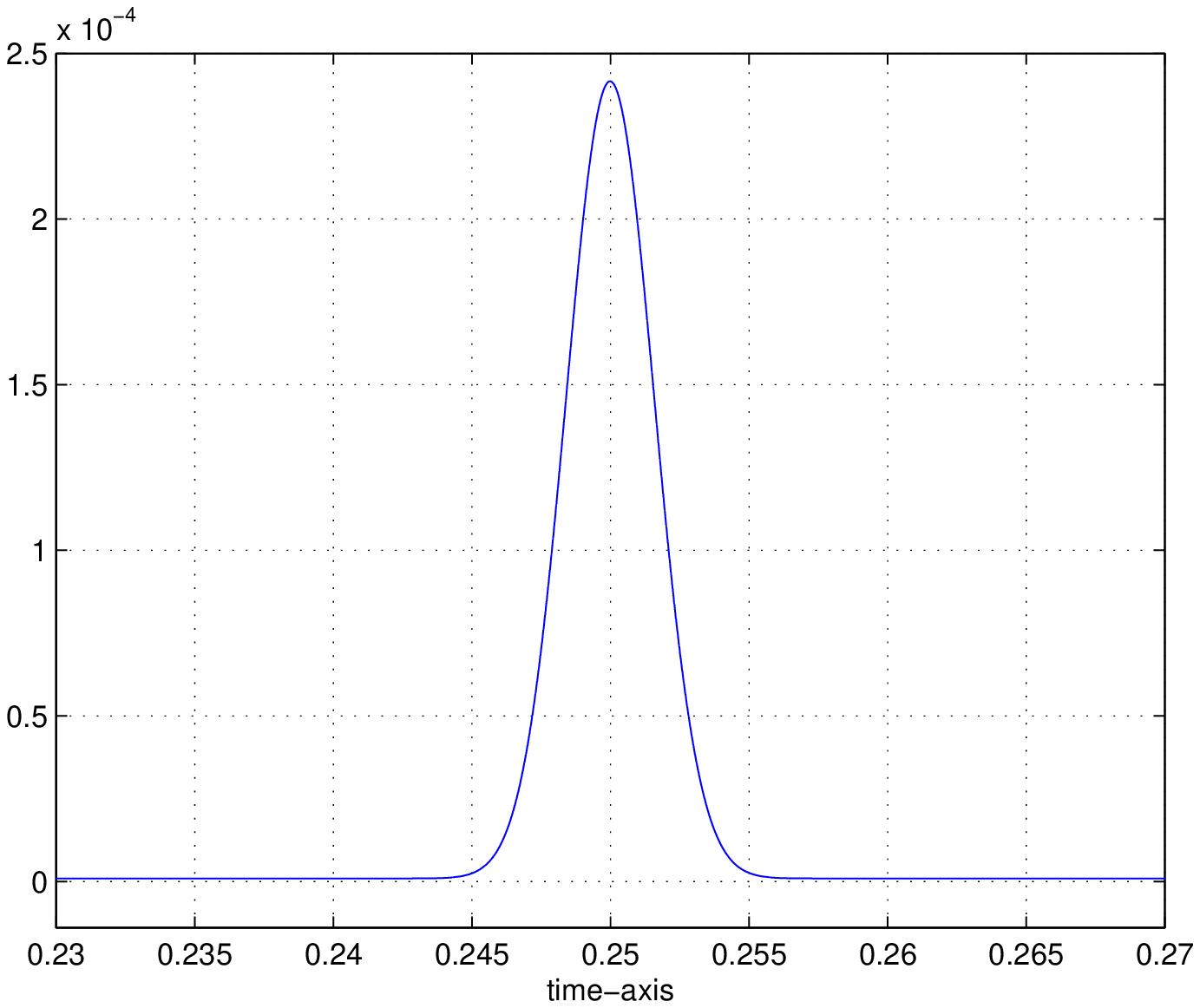}
\end{center}
\caption{{\sl Left:} $\ifourier{\exp{ \left(-\beta_*(\abs{\x},\omega)\right)}}$ defined by the thermo-viscous wave equation \req{thvwaveeq}
with $\tau_0=10^{-5}$, $c_0=1$ and fixed $\abs{\x}=\frac{1}{4}$.
{\sl Right:} Causal variant \req{cthvmodel} of the thermo-viscous wave equation with $\alpha_1=1$, $\tau_0=10^{-5}$, $c_0=1$ and fixed
$\abs{\x}=\frac{1}{4}$.
}
\label{fig:thviscous}
\end{figure}
\end{description}

\section{Conclusions}

In this paper we introduced the concept of an operator with causal domain of influence which guarantees a
finite wave front speed. As a consequence these models allow for a stable numerical implementation and thus are suitable for
photoacoustic imaging, where inversion techniques are required.
Based on this concept, we showed that an attenuated wave described by such an operator satisfies the standard
causality condition known as the \emph{Kramers-Kronig relation}. However theses relations are not sufficient to guarantee
that an attenuated wave has a finite wave front speed. This is a common misunderstanding in causality theory.

We also showed that attenuated waves satisfying the frequency power law and the Kramers-Kronig relation have
finite wave front speed only if $\gamma\in (0,1)$. An example of an equation where waves can propagate with infinite
wave front speed is the thermo-viscous wave equation.
Because of the lack of causality of standard models in the parameter range relevant for photoacoustic imaging, we
developed novel equations that satisfy our causality requirement and the desired attenuation properties.

For our causality analysis all equations were formulated as inhomogeneous equations with homogeneous initial conditions, but we showed that
if certain conditions are satisfied, then the attenuation problem can be formulated as a Cauchy problem with memory.

\section{Appendix: Nomenclature and elementary facts}

\begin{description}
\item{Real and Complex Numbers:} $\C$ denotes the space of complex numbers, $\R$ the space of reals. For a complex number $c=a+\imaginary b$
     $a=\Re{(c)}$, $b=\Im{(c)}$ denotes the real and imaginary parts, respectively.
\item{Differential Operators:} $\nabla$ denotes the gradient. $\nabla \cdot$ denotes divergence, and $\nabla^2$ denotes the Laplacian.
\item{Product:} When we write $\cdot$ between two functions, then it means a pointwise product, it can be a scaler product or if the functions
are vector valued an inner product. The product between a function and a number is not explicitly stated.
\item{Decomposition:} The decomposition of operators $\A$ and $\B$ is written as $\A\B$.
\item{Special functions:}
 The \emph{signum} function is defined by
\[
\sgn := \sgn (\x) := \frac{\x}{\abs{\x}}\;.
\]
In $\R^3$ it satisfies
\begin{equation}
\label{eq:der_sgn}
\nabla \cdot \sgn = \frac{2}{\abs{\x}}\;.
\end{equation}
The \emph{Heaviside} function
\[
 \Heavi := \Heavi(t) := \left\{ \begin{array}{rcl}
                      0 & \text{ for } & t < 0\\
                      1 & \text{ for } & t > 0\\
                     \end{array} \right.
\]
satisfies
\[
\Heavi := \frac{1}{2} (1+\sgn)\;.
\]
The Delta-distribution is the derivative of the Heaviside function at $0$ and is denoted by $\delta_t := \delta_t(t)$.
In our terminology $\delta_t$ denotes a \emph{one}-dimensional distribution.
The three dimensional Delta-distribution $\delta_\x$ is the product of three one-dimensional distribution $\delta_{x_i}$, $i=1,2,3$.
Moreover,
\begin{equation}
\label{eq:hatdelta}
\delta_{\x,t} := \delta_{\x,t}(\x,t) = \delta_\x \cdot \delta_t,
\end{equation}
is a four dimensional distribution in space and time.

\item{Properties related to functions:}
$\supp(g)$ denote the \emph{support} of the function $g$, that is the closure of the set of points, where $g$ does not vanish.
\item{Derivative with respect to radial components:}
We use the notation
\[
 r:=r(\x) = \abs{\x},
\]
and denote the derivative of a function $f$, which is only dependent on the radial component $\abs{\x}$, with respect to $r$
(i.e., with respect to $\abs{\x}$) by $\cdot'$.

Let $\beta = \beta(r)$, then it is also identified with the function $\beta = \beta (\abs{\x})$ and therefore
\[
 \nabla \beta = \frac{\x}{\abs{\x}} \beta'\;.
\]
\item{Convolutions:}
Three different types of convolutions are considered:
$*_t$ and $*_\omega$ denote \emph{convolutions} with respect to time and frequency, respectively.
Let $f$, $\hat{f}$, $g$ and $\hat{g}$ be functions defined on the real line with complex values. Then
\[
\begin{aligned}
 &f *_t g := \int_\R f(t-t')g(t') d t' , \quad\quad
 &\hat{f} *_\omega \hat{g}
        := \int_\R \hat{f}(\omega-\omega')\hat{g}(\omega') d \omega'.
\end{aligned}
\]
$*_{\x,t}$ denotes space--time convolution and is defined as follows:
Let $f,g$ be functions defined on the Euclidean space $\R^3$ with complex values, then
\[
f *_{\x,t} g := \int_{\R^3} \int_\R f(\x - \x',t-t')g(\x',t') d\x d t\;.
\]
\item{Fourier transform:} For more background we refer to~\cite{Lig64,Tit48,Pap62,Hoe03}.
All along this paper $\fourier{\cdot}$ is the Fourier Transformation with respect to $t$, and the inverse Fourier transform
$\ifourier{\cdot}$ is with respect to $\omega$.
In this paper we use the following definition of the \emph{Fourier transform} $\fourier{\cdot}$ and its inverse
$\ifourier{\cdot}$
\[
\fourier{f}(\omega) = \frac{1}{\sqrt{2\pi}} \int_\R \exp{(\imaginary \omega t)} f(t) d t, \quad
\ifourier{\hat{f}} (t) = \frac{1}{\sqrt{2\pi}} \int_\R \exp{(-\imaginary \omega t)} \hat{f}(\omega) d\omega\;.
\]
The Fourier transform and its inverse have the following properties:
\begin{enumerate}
\item \label{item:derF}
      \[
      \fourier{\timederivative{f}} (\omega) =  (-\imaginary \omega)\fourier{f}(\omega)\;.
      \]
\item \[
      \begin{aligned}
      \fourier{f \cdot g} & =\frac{1}{\sqrt{2\pi}} \fourier{f} *_\omega \fourier{g} \text{ and }\\
      \fourier{f} \cdot \fourier{g} &= \frac{1}{\sqrt{2\pi}} \fourier{f *_t g},\\
      \ifourier{\hat{f} \cdot \hat{g}} &= \frac{1}{\sqrt{2\pi}}\ifourier{\hat{f}} *_t \ifourier{\hat{g}} \text{ and }\\
      \ifourier{\hat{f}} \cdot \ifourier{\hat{g}} &= \frac{1}{\sqrt{2\pi}} \ifourier{\hat{f} *_\omega \hat{g}}\;.
      \end{aligned}
      \]
\item For $a \in \R$
\[
    \fourier{f(t-a)}(\omega) = \exp{(-i a \omega)} \cdot\fourier{f(t)}(\omega)
\]
\item \label{item:Fdelta}
    The Delta-distribution at $a \in \R$ satisfies
    \[
   \delta_t(t-a)=\frac{1}{\sqrt{2\pi}}\ifourier{\exp(\imaginary a \omega)}(t)\;.
\]
\item \label{item:even}
  Let $f$ be real and even, odd respectively, then $\fourier{f}$ is real and even, imaginary and odd, respectively.
\end{enumerate}
\item{The Hilbert transform for $L^2-$functions is defined by}
\[
 \hilbert{f} (t) = \frac{1}{\pi} \Xint-_\R \frac{f(s)}{t-s}ds\;,
\]
where $\Xint-_\R f(s) ds$ denotes the Cauchy principal value of $\int_\R f(s) ds$.

A more general definition of the  Hilbert transform can be found in~\cite{BelWoh66}.
The Hilbert transform satisfies
\begin{itemize}
 \item $\hilbert{\fourier{f}}(\omega) = -i\fourier{\sgn f}(\omega)$,
 \item $\hilbert{\hilbert{f}} = - f$.
\end{itemize}
From the first of these properties the Kramers-Kronig relation can be formally derived as follows. Since $f(t)$ is a
causal function if and only if $ f = \Heavi \cdot f$ and $\Heavi=(1+\sgn)/2$, it follows that
$\fourier{f} =  [\fourier{f} +\imaginary \hilbert{\fourier{f}}]/2 $, which is equivalent to
$\fourier{f} =  i\hilbert{\fourier{f}}$, i.e.
\[
      \Re(\fourier{f}) =  -\Im(\hilbert{\fourier{f}})  \ltext{and}
      \Im(\fourier{f}) =  \Re(\hilbert{\fourier{f}}) .
\]
\item{The inverse Laplace transform} of $f$ is defined by
\[
\ilaplace{f}(t) =
\left\{ \begin{array}{ccl}
0 &\text{ for }& t < 0,\\
\frac{1}{2 \pi i} \int_{\gamma - i\infty}^{\gamma + \imaginary \infty} \exp{(st)} f(s)ds, &\text{ for }& t > 0,
\end{array}\right.
\]
where $\gamma$ is appropriately chosen.

The inverse Laplace transform satisfies (see e.g. \cite{Heu91}) that
\begin{equation}
\label{eq:pr1}
\ilaplace{h(s-a)}(t) = \exp{(a t)} \ilaplace{h(s)}(t) \text{ for all }
a, t \in \R
\end{equation}
and
\begin{equation}
 \label{eq:pr2}
\ilaplace{s^{-r}}(t) = \frac{H(t) t^{r-1}}{\Gamma(r)}\qquad (r>0)\;.
\end{equation}
\end{description}

\section*{Acknowledgement}
This work has been supported by the Austrian Science Fund (FWF)
within the national research network Photo\-acoustic Imaging in Biology and Medicine, project S10505-N20.
The stay of XB in Innsbruck was partly supported by the ``Frankreich-Schwerpunkt'' of the University of Innsbruck.

\def\cprime{$'$} \def\cprime{$'$} \def\cprime{$'$}

\end{document}